\documentclass[a4paper,12pt]{article}
\pdfoutput=1 

\usepackage{jcappub} 

\usepackage[T1]{fontenc} 

\usepackage{amsmath}
\usepackage{color}
\usepackage[utf8]{inputenc}
\usepackage[normalem]{ulem}
\usepackage{graphicx}
\usepackage{subcaption}
\usepackage{cleveref}
\usepackage{amsthm}

\title{\boldmath Thermodynamics of dyonic black holes in non-linear electrodynamics}

\author[a]{Lewis Croney,}
\author[a,b]{Ruth Gregory,}
\author[a,1]{Carlos J. Ramírez-Valdez\note{Corresponding author.}}

\affiliation[a]{Department of Physics, King’s College London, University of London, Strand, London, WC2R 2LS, UK}
\affiliation[b]{Perimeter Institute, 31 Caroline Street North, Waterloo, ON, N2L 2Y5, Canada}

\emailAdd{lewis.croney@kcl.ac.uk}
\emailAdd{ruth.gregory@kcl.ac.uk}
\emailAdd{jonathan.ramirez@tec.mx}

\abstract{We investigate dyonic black holes in a weak field expansion of non-linear electrodynamics.
The breadth of parameter space permits a rich thermodynamic structure, additional turning points and intricate phase phenomena.
Energy conditions are employed to ensure the physical viability of
solutions.
Analytic special cases illustrate novel properties of black holes in non-linear electrodynamics, including modified extremal limit behaviour.
Numerical solutions offer the most elaborate thermodynamic landscape, culminating in up to five turning points, and multiple reentrant phase transitions.}

\begin{document}
\maketitle
\flushbottom

\section{Introduction}

The study of black hole thermodynamics is a vibrant area of research in theoretical physics \cite{bekenstein1973,bekenstein1974,israel1986-thirdlaw,wald2001}, with significant implications for our understanding of gravitational systems from a distinctive perspective, potentially shedding light on intrinsic quantum behaviour \cite{hawking1975}. One particular aspect that has been recently illuminated is the exploration of the extended thermodynamic phase space of black hole solutions \cite{kastor2009,
kubiznak2012,mann2017}, or indeed the thermodynamics of more complicated black hole spacetimes such as accelerating black holes \cite{Appels:2016uha,Anabalon:2018qfv}. With input from a wider array of matter content, these extended black hole solutions can allow rich phase phenomenology, such as reentrant phase transitions \cite{Frassino:2014pha}, swallowtails \cite{chamblin1999-april} and even snapping swallowtails \cite{Abbasvandi:2018vsh, Gregory:2019dtq}!
These interesting phenomena can occur particularly when there are non-linear interactions in the electromagnetic sector; here regular black hole solutions have also been developed, circumventing the singularity problem by introducing a non-singular, well-behaved core \cite{bardeen1973, ayon1999}.

The study of non-linear electrodynamics (NLED) has undergone significant advancements since its inception with the seminal work of Born and Infeld \cite{born1934}, who sought to resolve the divergent self-energy of point charges in classical electrodynamics. More recently, NLED has found applications in string theory \cite{gibbons1998}, gauge/gravity duality \cite{hartnoll2009} and black hole thermodynamics \cite{Okyay:2021nnh}. The application of NLED in anti-de Sitter (AdS) space has provided crucial insights, for instance into holographic superconductors and the AdS/CFT correspondence \cite{hartnoll2008, Zhao:2012cn}

The thermodynamic properties of NLED in the presence of strong gravitational fields has been a subject of intense research, particularly in the context of black hole solutions that exhibit elaborate phase structures, including Van der Waals-like behaviour \cite{kubiznak2012}.
Recent investigations into thermodynamics of black holes in NLED have revealed deep connections between non-linear electromagnetic fields and quantum gravitational effects. Extending their work in \cite{Kubiznak:2012wp} to non-linear theories, Gunasekaran, Mann, and Kubiznak \cite{kubiznak2012} showed that NLED corrections can significantly alter Reissner-Nordström-AdS (RNAdS) black hole thermodynamics, leading to new critical phenomena in the extended phase space. 
Reentrant phase transitions have also been observed with black holes in NLED theories \cite{Dehyadegari:2017hvd}, which are known to occur in quantum systems such as Bose-Einstein condensates \cite{PhysRevLett.93.160402}.

In linear Maxwell theory, the black hole solutions are symmetric under an exchange of electric and magnetic charges. However, in more intricate theories, such as those with additional fields \cite{Gibbons:1987ps}, or in holography \cite{hartnoll2009}, this symmetry can be broken. Exploring black holes carrying both electric and magnetic charges, so-called \textit{dyonic} configurations, opens the door to a landscape of distinctive behaviour, particularly in the context of NLED where this symmetry is also broken \cite{Bronnikov:2017xrt}.

Some works have considered black holes in \textit{exact} NLED theories, such as Born-Infeld \cite{Demianski:1986wx, Wiltshire:1988uq, kubiznak2012}. 
More recently, the approach of expanding the NLED theory in electromagnetic invariants $F= F_{\mu\nu}F^{\mu\nu}$ and $G=F_{\mu\nu}\Tilde{F}^{\mu\nu}$ has been extensively studied \cite{yajima2001, Magos:2020ykt, Karakasis:2022xzm, kruglov2024}.
Many works have restricted attention to either electrically or magnetically charged black holes, or asymptotically flat space.
However, no such study has explored the thermodynamics of the full parameter space of dyonic black holes in AdS, with independent couplings for the lowest order non-linear terms $F^2$ and $G^2$.

In this work, we address this gap in the literature,
presenting a comprehensive analysis of the thermodynamics of black holes with weak non-linear electromagnetic fields. Our work bridges recent advances in extended phase space thermodynamics \cite{kastor2009, mann2017} with the unique features of NLED, offering insights into the stability and criticality of these systems.
We begin by fixing the NLED action under consideration, and discussing its generality as a weak-field expansion.
In deriving the black hole solutions, we can compare new attributes against the standard RNAdS case. 
After reviewing the usual story of phase transitions in the linear theory, we determine the thermodynamics of the non-linear system. Working in asymptotically AdS spacetime, the extended Smarr relation and first law of thermodynamics are derived.
The approach adopted in this work involves studying the phase transition structure of the gravitational solution, which in turn entails analysing the turning points of the corresponding thermodynamic potential, the Gibbs free energy \cite{hawking-page1983,chamblin1999,chamblin1999-april,kruglov2024}.
Working in an analytic perturbative expansion, we start to elucidate the complex structure of the non-linear effects, with additional turning points at play.
To reveal the full depth of the interplay of electric and magnetic charges, with non-linearities in the theory, we proceed numerically, finding a maximum number of five turning points. We discuss the role of heat capacity in determining thermodynamic stability of NLED black holes, and we explore the possible phase transitions, finding an example of a multiple reentrant phase transition between three branches.

\section{Dyonic black hole solution} \label{SolutionOverallSec}

The system under consideration in this work has the following action,
\begin{align}\label{action}
\mathcal{S}=\frac{1}{16\pi}\int_{\mathcal{M}^4} d^4x \sqrt{-g} (R-2\Lambda-F+a F^2+b G^2),
\end{align}
where $R$ is the Ricci scalar, $\Lambda$ is the cosmological constant, and $F$ and $G$ are electromagnetic invariants defined previously as $F:= F_{\mu\nu}F^{\mu\nu}$ and $G:=F_{\mu\nu}\Tilde{F}^{\mu\nu}$. 
There is the usual Maxwell linear $F$ term, but there is no need to have a term linear in $G$, since it is topological and would not contribute to the equations of motion.

Many physically interesting theories of non-linear electrodynamics can be written in the form \eqref{action}, when expanded in a weak field approximation to second order. 
For example, Euler-Heisenberg theory arises as a one-loop QED correction to Maxwell electrodynamics \cite{heisenberg1936}, which in its weak-field approximation gives a theory of the form \eqref{action}.
In terms of the electric and magnetic fields the electromagnetic part of the Euler-Heisenberg Lagrangian can be expressed in flat space as \cite{heisenberg1936},
\begin{align} \label{EulerHeisenbergLagrangian}
\mathcal{L_{\textrm{EM}}}=\frac{1}{4\pi}  \left[\frac{1}{2}(\mathbf{E}^2-\mathbf{B}^2)+\frac{2\alpha^2}{45 m_e^4}\left[(\mathbf{E}^2-\mathbf{B}^2)^2+7(\mathbf{E}\cdot \mathbf{B})^2\right]\right],
\end{align}  
so that the specific coupling constants are $a=2\alpha^2/45 m_e^4$ and $b=7\alpha^2/90 m_e^4$, for fine structure constant $\alpha$ and electron mass $m_e$.

Furthermore, Born-Infeld theory \cite{born1934} was originally constructed to make the self energy of charged particles finite. It has the matter Lagrangian,
\begin{align}
    \mathcal{L}_{EM} = \frac{T}{4\pi\sqrt{-g}}\left(\sqrt{-g} - \sqrt{-\text{det}(g_{\mu \nu} + T^{-1/2}F_{\mu\nu} )}\right),
\end{align}
where $g_{\mu \nu}$ is the full spacetime metric, and $T$ is a positive coupling constant. In its weak field form, Born-Infeld becomes \cite{Sorokin:2021tge},
\begin{align} \label{BornInfeldLagrangian}
\mathcal{L_{\textrm{EM}}}=\frac{1}{16\pi}  \left[-F + \frac{1}{8T}(F^2+G^2)\right],
\end{align} 
taking the form of \eqref{action}, with $a = b = \frac{1}{8T}$.

However, the action in \eqref{action} is more general than just the examples provided.
Many theories of non-linear electrodynamics fall under the category of $\mathcal{L}(F,G)$ theories \cite{Bokulic:2021dtz}, where the electromagnetic sector is a function of $F$ and $G$. Provided $\mathcal{L}(F,G)$ is analytic at $(F, G) = (0,0)$, it can be expanded in a weak-field approximation (for small $F,G$),
\begin{align} \label{LFGexpansion}
    \mathcal{L}(F, G) =a_{00} + a_{10} \, F + a_{01} \,G + a_{20} \,F^2 + a_{11} \, FG + a_{02} \, G^2 + ... \; .
\end{align}
Without loss of generality, we can set $a_{00} = 0 $ to absorb it into the cosmological constant, $a_{10} = -1$ to match Maxwell at low energy, and $a_{01} = 0$ as a $G$ term is topological.
If we further require parity $P$ invariance, and time-reversal $T$ invariance, where,
\begin{equation}
\begin{array}{r@{\;}l}
P: & 
\left\{
\begin{aligned}
F &\rightarrow F \\
G &\rightarrow -G
\end{aligned}
\right.
\end{array}
\quad , \quad
\begin{array}{r@{\;}l}
T: & 
\left\{
\begin{aligned}
F &\rightarrow F \\
G &\rightarrow -G
\end{aligned}
\right.
\end{array}
\quad ,
\end{equation} 
then the $FG$ term is excluded, and we can set $a_{11} = 0$. Then \eqref{LFGexpansion} is of the form \eqref{action}, which is hence the most general second order parity and time-reversal invariant weak-field expansion of such a non-linear electromagnetic sector.

It is worth noting that not all $\mathcal{L}(F, G)$ theories fall in this category. For example, the recent ModMax theory \cite{Bandos:2020jsw},
\begin{align} \label{ModMaxLag}
    \mathcal{L} = \frac{1}{16\pi}\left[-(\cosh\gamma) \, F + (\sinh\gamma) \sqrt{F^2 + G^2}\right],
\end{align}
with coupling constant $\gamma$, is not analytic at $(F, G) = (0,0)$, and hence cannot be expanded in a weak-field approximation as in \eqref{action}.

Rather than focusing on a particular example, in this work we will consider the general form \eqref{action}, and treat $a$, $b$ as independent coupling constants. However, we will restrict our consideration to $a, b \geq 0$, as we outline in Section \ref{DECVioSec}, where we find that energy conditions also limit the magnitudes of $a$ and $b$. 

In spite of the fact that our theory \eqref{action} is only a second-order weak-field expansion, we nonetheless take the perspective that we should fix our theory, and fully explore its consequences, even if it is only an approximation of some more complete theory, in the same way that Einstein-Hilbert is only the first term in an EFT expansion of gravity.

The corresponding equations of motion for the action in \cref{action} are given by,
\begin{align}
    R_{\mu\nu}-\frac{1}{2}g_{\mu\nu}R+g_{\mu\nu}\Lambda &= 8\pi T_{\mu\nu},\label{eom-grav}\\
    \nabla^\mu P_{\mu\nu}&=0,\label{eom-em}
\end{align}
where, the energy momentum tensor $T_{\mu\nu}$, and the auxiliary tensor $P_{\mu\nu}$ are given by,
\begin{align}
    T_{\mu\nu} &= \frac{1}{4\pi}\left[\left(1-2aF\right)F_{\mu\alpha}F^{\hspace{0.4 em}\alpha}_\nu-\frac{1}{4}g_{\mu\nu}(F-aF^2+bG^2)\right], \label{EngMomTensor}\\
    P_{\mu\nu} &= F_{\mu\nu}-2aFF_{\mu\nu}-2bG\Tilde{F}_{\mu\nu}.\label{Pdef}
\end{align}
Without the non-linear $a,b$ terms, the electromagnetic equations would take their vacuum form $\nabla_{\mu} F^{\mu \nu} =0$. We could view the additional terms in \eqref{eom-em}, \eqref{Pdef} as providing an additional current, modifying the equations to the form $\nabla_{\mu} F^{\mu \nu} = \mu_0 J^{\nu}$. With an effective current, we should not generally expect electromagnetic duality in $p \leftrightarrow q$ that exists in the dyonic Reissner-Nordstr\"om solution. 

Many works on black holes in NLED tend to take the ``metric engineering'' approach \cite{Bronnikov:2000vy, dePaula:2025kif}; first identify the feature one plans to observe, and then determine the non-linear theories that give rise to the feature. Here we take a more direct approach: we begin with a physically-motivated fixed NLED theory, then derive the black hole solutions and investigate interesting features they display.
If we look for a static, spherically symmetric solution, then since $T^t_t = T^r_r$ it can be proven that the metric takes the form \cite{Bowcock:2000cq, yajima2001},
\begin{equation}\label{metric}
ds^2 = - f(r) dt^2 + \frac{1}{f(r)}dr^2 + r^2 \left(d\theta^2 + \sin^2 \theta d\varphi^2\right). 
\end{equation}
(Note, the symmetry assumption guarantees that the only two independent components of $F_{\mu \nu}$ are $F_{rt}$ and $F_{\theta \varphi}$, and hence the expressions for $T^t_t$ and $T^r_r$ are identical.)
The electromagnetic field equations \eqref{eom-em}, along with the Bianchi identity $\nabla^{\mu} \tilde{F}_{\mu \nu} =0$, then immediately fix two components,
\begin{align}
    P_{rt}&=-P_{tr}=\frac{q}{r^2},\label{Pans}\\
    F_{\theta\varphi}&=-F_{\varphi\theta}=p\sin{\theta},\label{Fans}
\end{align}
where $q,p$ are constants that reduce to the electric and magnetic charge of the black hole in the linear case. 
The remaining components to determine are $P_{\theta\varphi}$ and $F_{rt}$. Only $F_{rt}$ will be significant for solving the Einstein equations, since the energy-momentum tensor is expressed in terms of $F_{\mu \nu}$ in equation \eqref{EngMomTensor}. 
The electrodynamic contractions are,
\begin{align}
    F &= F_{\mu\nu}F^{\mu\nu} = \frac{2p^2}{r^4} - 2 F_{rt}^2 \label{Fcontraction}, \\
    G &=F_{\mu\nu}\Tilde{F}^{\mu\nu} = \frac{4pF_{rt}}{r^2}. \label{Gcontraction}
\end{align}
The component $F_{rt}$ can then be determined by taking the $(r, t)$ component of equation \eqref{Pdef}, giving the cubic,
\begin{equation}
    4 a F_{rt}^3 + \left(1 + \frac{4p^2}{r^4}(2b-a) \right) F_{rt} - \frac{q}{r^2} = 0.\label{Fcubic}
\end{equation}
For large $r$, we see that $F_{rt} = q/r^2 + O(r^{-6})$, confirming that $q$ is indeed the asymptotic electric charge of the spacetime.
In general, the cubic for $F_{rt}$ could have at most three real roots. However, consistency with $F_{rt} \approx q/r^2$ as $r \rightarrow \infty$ dictates the unique physical solution.
For example, in the Euler-Heisenberg and Born-Infeld cases, since $a > 0$ and $2b - a > 0$, we are guaranteed only one real solution to \cref{Fcubic},
\begin{equation}
    F_{rt} = (u_+)^{1/3} + (u_-)^{1/3}, \quad u_{\pm} \equiv \frac{q}{8 a r^2} \left(1 \pm \sqrt{1 + \frac{(r^4 + 4(2b-a)p^2)^3}{27 a q^2 r^8}}\right).
\end{equation}
Other cases, in particular where $2b-a < 0$, may have more than one solution for some values of $r$, and hence more care is required to identify the physical solution.

This cubic will complicate the analysis of the black hole solutions below. We will express the solution generally in terms of $F_{rt}$, with the understanding that the cubic solution could be employed. Nonetheless, we will explore two special cases for which the cubic has immediate solutions, when $a = 0$, and when $q = 0$, to gain an analytic appreciation of new features the non-linear terms present. Later we will employ numerical methods to appreciate the full richness of the available parameter space.

Once we understand $F_{rt}$, we have all the ingredients to determine the metric.
Taking the $(t, t)$ component of the Einstein equations \eqref{eom-grav}, we find that $f(r)$ satisfies the equation,
\begin{equation}
    \frac{(r f(r))' - 1}{r^2} + \Lambda = 8 \pi T^t_t = \frac{1}{2} \left(1 + \frac{4p^2}{r^4}(2b-a) \right) F_{rt}^2 - \frac{3q}{2r^2} F_{rt} + \frac{2a p^4}{r^8} - \frac{p^2}{r^4},
\end{equation}
where the cubic equation \eqref{Fcubic} has been applied to reduce the powers of $F_{rt}$.
Rearranging and integrating, we find the schematic form,
\begin{equation} \label{frfull}
    f(r) = 1 - \frac{2m}{r} - \frac{\Lambda r^2}{3} + \frac{p^2}{r^2} -\frac{2ap^4}{5r^6}+ \frac{1}{2r} \int^r dr' \left[\left(r'^2 + \frac{4p^2}{r'^2}(2b-a) \right) F_{rt}^2 - 3q F_{rt} \right],
\end{equation}
where $m$ is a constant of integration usually associated with the black hole ADM mass.

\section{Analytic study: the magnetic black hole}
\label{Analyticq0Sec}

Performing the integral in \eqref{frfull} is messy in general, since $F_{rt}$ is the root of the cubic \eqref{Fcubic}.
In this section, we develop a simple example that illuminates some of the interesting attributes that NLED black holes have to offer.

\begin{figure}[t]
    \centering
        
    \begin{minipage}{0.47\textwidth}
        \includegraphics[width=\textwidth]{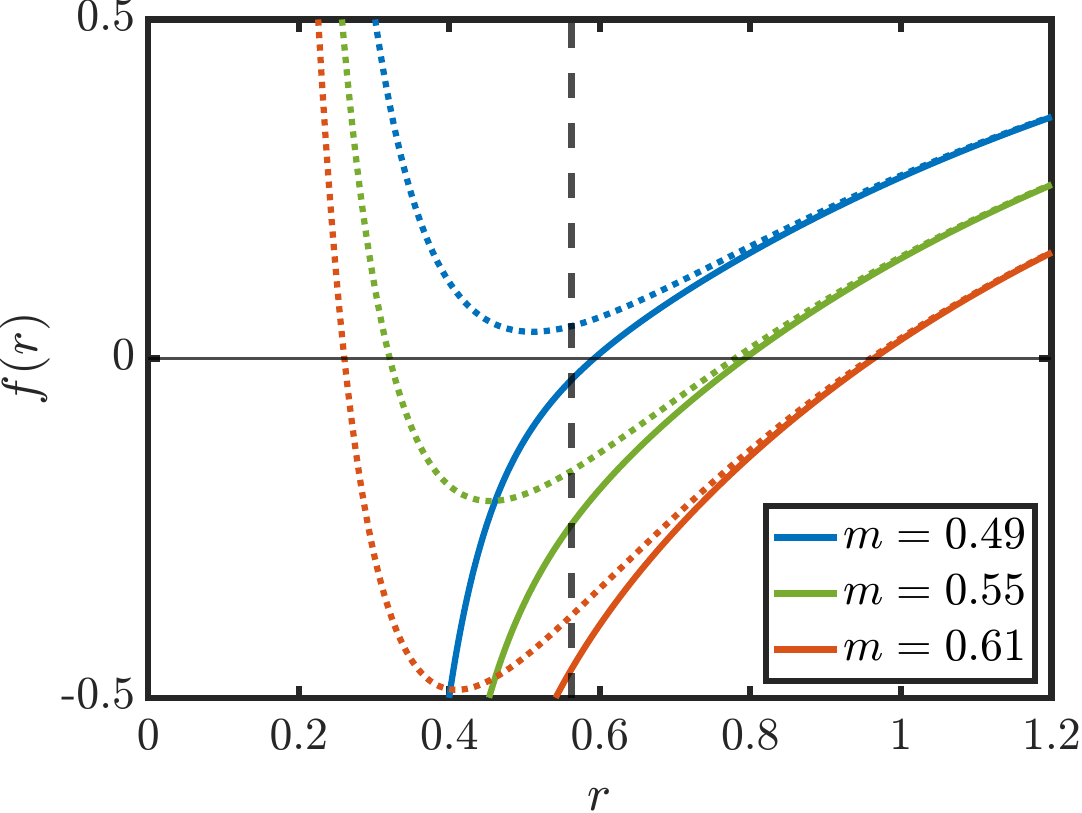}
    \end{minipage}
    \hspace{0.4cm}
    \begin{minipage}{0.47\textwidth}
        \includegraphics[width=\textwidth]{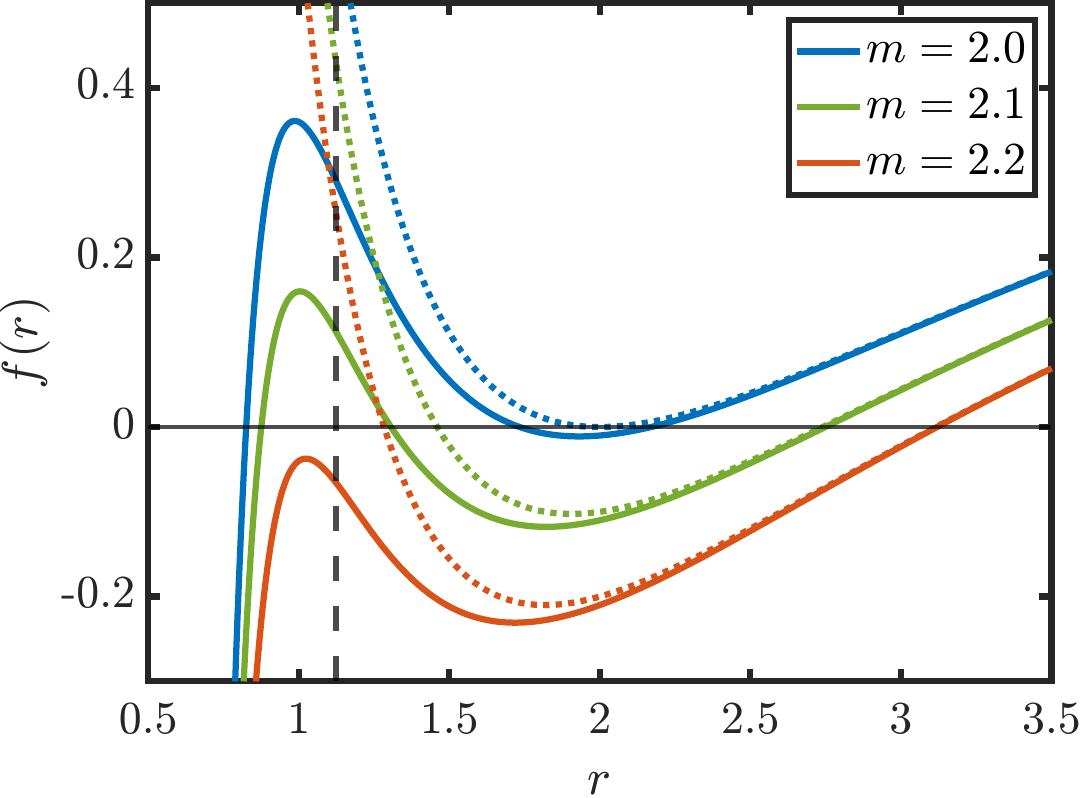}
    \end{minipage}
    
    \caption{Plots of $f(r)$ against $r$ for $\Lambda = 0$, $q = 0$, for different masses $m$ and magnetic charge $p$. The left plot is for $p = 0.5$ and the right is for $p = 2$.
    The solid curves are the non-linear results with $a = 0.1$, and the dotted curves are their linear counterparts with $a = 0$.
    The horizontal thin line marks $f(r) = 0$, where the horizons occur. The vertical dashed line marks the threshold for the DEC violation discussed in Section \ref{DECVioSec}, so each solution shown conserves DEC outside the horizon.
    In the left plot, the linear theory displays an extremal limit as $m$ is varied, and yet the non-linear $f(r)$ are monotonic. In the right plot, it appears that the horizon jumps as $m$ is decreased, although to a region with DEC violation. We investigate these features further in Section \ref{ExtremalJumpSec}.
    For large $r$, all non-linear curves approach their linear counterparts, which have $f(r) \rightarrow 1$. }
    \label{q0f(r)Ex}
\end{figure}

The non-linearities in the action \eqref{action} can give rise to a significant modification in the metric function $f(r)$ compared to the linear case.
For sufficiently large $r$, since $F_{rt} \sim q/r^2$, the new terms in $f(r)$ will be subdominant, and we will recover $f(r) \sim f(r)_{\text{RN(A)dS}}$. However, for small $r$, additional contributions may modify the picture.
To illustrate this, we consider the simple case $q = 0$, where the physical solution of the cubic \eqref{Fcubic} would be $F_{rt} \equiv 0$, and the integral in \eqref{frfull} vanishes.
In that case, we find,
\begin{align} \label{f(r)q0eq}
    f(r) = 1 - \frac{2m}{r} - \frac{\Lambda r^2}{3} + \frac{p^2}{r^2} -\frac{2ap^4}{5r^6} = f(r)_{\text{RN(A)dS}} - \frac{2ap^4}{5r^6}.
\end{align} 
Examples of $f(r)$ for this analytic example are shown in Figure \ref{q0f(r)Ex}.
Since $F_{rt} = 0$, equation \eqref{Gcontraction} shows that $G = 0$, and hence $b$ plays no role in this case.
In RN(A)dS, for small $r$, the dominant divergence is from a positive $r^{-2}$ term, but now this gives way to a negative $r^{-6}$ term. In the linear theory, there is an extremal limit when $m = p$, with a naked singularity for $m < p$. 
Furthermore, the domination of this negative final term at small $r$ guarantees a horizon \textit{for all} $m$.  However, energy conditions prevent this new term from drastically overpowering the others, and will disallow some regions of parameter space, as we come to discuss below.
It is worth noting that instead setting $p = 0$ does not lead to the same simplification for $f(r)$, which is an example of the explicit breaking of the $p \leftrightarrow q$ electromagnetic duality that is present in vacuum linear electromagnetism.

\subsection{Energy conditions} \label{DECVioSec}

Energy conditions can be imposed on a theory with matter content to ensure its physical plausibility. Violations of these conditions can lead to spacetimes with exotic behaviour, such as traversable wormholes \cite{Morris:1988tu} and warp drives \cite{Alcubierre:1994tu}. 

To ensure that our theory \eqref{action}, obtained by restricting to second order in the fields, offers plausible results, we can use energy conditions to restrict the viable couplings $a$, $b$. Energy conditions have been considered previously in theories of non-linear electrodynamics \cite{Bokulic:2021dtz, Russo:2024xnh}, but we derive them here in the context of our black hole solution.

These conditions can be expressed quite simply by working in an orthonormal frame, following the treatment of \cite{Poisson:2009pwt},
\begin{align}
    e^{\alpha}_{\hat{t}} = \frac{1}{\sqrt{f(r)}} \delta^{\alpha}_{t}, \quad e^{\alpha}_{\hat{r}} = \sqrt{f(r)} \delta^{\alpha}_{r}, \quad e^{\alpha}_{\hat{\theta}} = \frac{1}{r}\delta^{\alpha}_{\theta}, \quad e^{\alpha}_{\hat{\varphi}} = \frac{1}{r \sin\theta} \delta^{\alpha}_{\varphi},
\end{align}
so that $g_{\alpha \beta} \, e^{\alpha}_{\hat{\mu}} \, e^{\beta}_{ \hat{\nu}} = \eta_{\hat{\mu} \hat{\nu}}$.
We can convert the energy-momentum components \eqref{EngMomTensor} into this frame,
\begin{equation}
    \begin{aligned}
        \rho &\equiv T_{\hat{t} \hat{t}} = \frac{1}{8\pi} \left(6a F_{rt}^4 + \left(1 + \frac{4(2b-a)p^2}{r^4}\right)F_{rt}^2 + \frac{p^2}{r^4} - \frac{2ap^4}{r^8}\right), \\
        p_r &\equiv T_{\hat{r} \hat{r}} = -\rho, \\
        p_{\theta} &\equiv T_{\hat{\theta} \hat{\theta}} = \frac{1}{8\pi} \left(2a F_{rt}^4 + \left(1 - \frac{4(2b-a)p^2}{r^4}\right)F_{rt}^2 + \frac{p^2}{r^4} - \frac{6ap^4}{r^8}\right), \\
        p_{\varphi} &\equiv  T_{\hat{\varphi} \hat{\varphi}} = p_{\theta},
    \end{aligned}
\end{equation}
where $p_i$ are pressures, not to be confused with the magnetic charge $p$. Note that we do not include the cosmological constant contribution in the energy-momentum tensor, instead treating it as part of the background geometry.
In terms of these quantities, the energy conditions read as \cite{Poisson:2009pwt},
\begin{equation}
    \begin{aligned}
        \text{Null (NEC): }& \; \rho + p_{i} \geq 0, \\
        \text{Weak (WEC): }& \; \rho \geq 0, \quad \rho + p_{i} \geq 0, \\
        \text{Dominant (DEC): }& \; \rho \geq 0, \quad \rho \geq |p_{i}|, \\
        \text{Strong (SEC): }& \; \rho + p_r + p_{\theta} + p_{\varphi} \geq 0, \quad \rho + p_{i} \geq 0. \\
    \end{aligned}
\end{equation}

Since we only have two independent energy-momentum components, related to $\rho$ and $p_{\theta}$, these energy conditions collapse to,
\begin{equation}
    \begin{aligned}
        \text{NEC: }& \; \rho + p_{\theta} \geq 0, \\
        \text{WEC: }& \; \rho \geq 0, \quad \rho + p_{\theta} \geq 0, \\
        \text{DEC: }& \; \rho \geq 0, \quad \rho \geq |p_{\theta}|, \\
        \text{SEC: }& \; p_{\theta} \geq 0, \quad \rho + p_{\theta} \geq 0. \\
    \end{aligned}
\end{equation}

We will choose to respect the dominant energy condition here, automatically guaranteeing the weak and null energy conditions. Since there are many physically relevant theories that violate the strong energy condition, such as inflation \cite{Visser:1999de} and domain walls \cite{Ipser:1983db}, we will permit such violations here. 
However, there has been a significant body of literature connecting causality, a more physically-motivated principle, to energy conditions \cite{Russo:2024xnh, Tomizawa:2023vir, Abe:2025vdj}. In pure Maxwell theory, electromagnetic perturbations propagate along the null cones of the spacetime metric, but in NLED the non-linearities cause these perturbations to follow null cones of modified effective metrics \cite{Novello:1999pg}, which alters the causal structure.
The authors of \cite{Schellstede:2016zue} consider causality to be violated when the propagation of these perturbations is superluminal with respect to the \textit{spacetime} metric, and they derive constraints on NLED theories to prevent this superluminal propagation.
Other authors have suggested that superluminal velocities may not be sufficient to induce genuine causality violations in the sense of closed timelike curves \cite{Drummond:1979pp, Shore:1995fz, Goon:2016une}.
Nonetheless, with the definition and constraints of \cite{Schellstede:2016zue}, the authors of \cite{Russo:2024xnh} have shown that the absence of superluminal propagation implies the SEC. Hence, for a choice of parameters that violates the SEC, the theory \eqref{action} allows for superluminal propagation, and hence may violate causality. 
The implications for these solutions are therefore a potential concern, with superluminal propagation perhaps pointing to an instability \cite{Aharonov:1969vu, Moreno:2002gg}. These issues warrant further investigation in the context of these solutions, but we leave this to future work.

We will only require these DEC inequalities to be satisfied outside the horizon of the black hole, since only this region is accessible in the coordinate system of \eqref{metric}.

As such, energy conditions enforce the following inequalities,
\begin{align}
    \rho \geq 0 & \quad \Rightarrow \quad 6a F_{rt}^4 + \left(1 + \frac{4(2b-a)p^2}{r^4}\right)F_{rt}^2 + \frac{p^2}{r^4} - \frac{2ap^4}{r^8} \geq 0, \label{rho_positive} \\
    \rho + p_{\theta} \geq 0 & \quad \Rightarrow \quad 8a F_{rt}^4 + 2F_{rt}^2 + \frac{2p^2}{r^4} - \frac{8ap^4}{r^8} \geq 0, \label{rho_plus_p_positive} \\
    \rho - p_{\theta} \geq 0 & \quad \Rightarrow \quad 4a F_{rt}^4 + \frac{8(2b-a)p^2}{r^4}F_{rt}^2  + \frac{4ap^4}{r^8} \geq 0 \label{rho_minus_p_positive}.
\end{align}
Using the expressions for $F, G$ in \eqref{Fcontraction}, \eqref{Gcontraction}, the latter constraint \eqref{rho_minus_p_positive} can be factorised into the form,
\begin{align} \label{abconstraint}
     aF^2 + bG^2 \geq 0.
\end{align} 
We can trivially satisfy this constraint by restricting our attention to $a,b \geq 0$. The weak field expansion of both the Euler-Heisenberg \eqref{EulerHeisenbergLagrangian} and Born-Infeld \eqref{BornInfeldLagrangian} theories fall within this category. Since our parameter space in $(p, q, a, b, m, \Lambda)$ is already sufficiently rich, we leave consideration of $a < 0$, $b < 0$ to future study, where it may still be possible to evade this constraint \eqref{abconstraint}. Although, it has been argued that $a, b \geq 0$ is necessary to respect causality \cite{Abe:2025vdj}.

Having fixed the sign of $a,b$, we now turn to their magnitudes. The other conditions \eqref{rho_positive} and \eqref{rho_plus_p_positive} can place non-trivial constraints on the allowed parameter space.

We can probe this in the simple case of $q = 0$, where the remaining constraints become,
\begin{align}
    \rho \geq 0 & \quad \Rightarrow \quad \frac{p^2}{r^4} - \frac{2ap^4}{r^8} \geq 0, \\
    \rho + p_{\theta} \geq 0 & \quad \Rightarrow \quad \frac{2p^2}{r^4} - \frac{8ap^4}{r^8} \geq 0,  \end{align}
with the latter placing the strongest constraint,
$r_+^4 \geq 4ap^2$.

\begin{figure}[ht]
\centering
\includegraphics[width=0.6\textwidth]{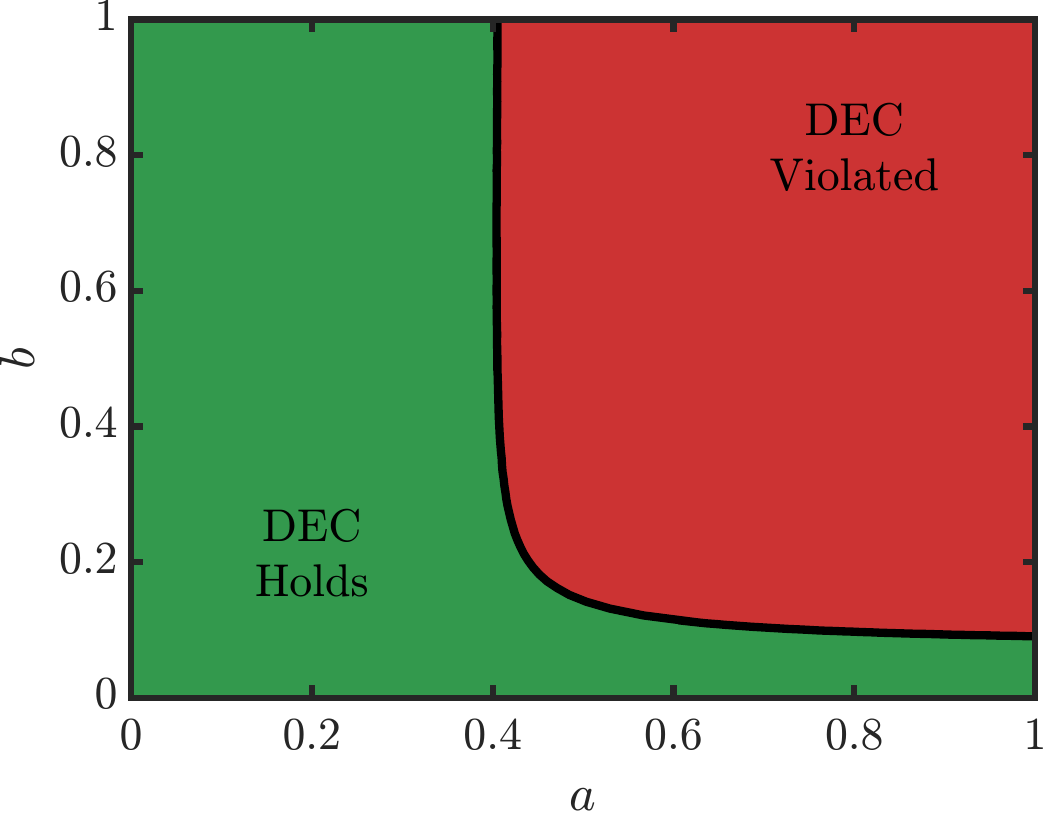}
\caption{A plot showing parameters $(a,b)$ that satisfy and violate DEC, for $p = 0.1$, $q = 0.3$, $m = 0.38$ and $\Lambda = 0$. If the event horizon $r_+$ is \textit{less} than the radius from the DEC constraint, then DEC is violated outside the horizon, and the choice of parameters $(a,b)$ are excluded.}
\label{DECContour}
 \end{figure}

While analytic treatment is challenging when moving away from the particular case $q = 0$, since $F_{rt}$ generally solves the cubic \eqref{Fcubic}, a numerical analysis confirms that the picture broadly persists. 
For example, in Figure \ref{DECContour}, we determine exclusion bounds for $a,b$  numerically for fixed $p,q \neq 0$. Generally we see that larger $a$ and $b$ tend to be more likely to lead to DEC violation, constraining the possible parameter space for physically viable black hole configurations.

There are also families of solutions where DEC violation does not occur. For example if $a = 0$ and $b\geq0$, or separately if $p = 0$ and $a \geq 0$, then we see that \eqref{rho_positive} and \eqref{rho_plus_p_positive} are always satisfied for $r>0$. Of course the linear theory also provides such an example. We will avoid cases with any DEC violation outside the black hole event horizon in any results we report below, with DEC violation thresholds shown on any relevant plots.

\subsection{Extremal limits and horizons} \label{ExtremalJumpSec}

In the linear theory, the black hole solutions always come with an extremal limit, occurring for $m = p$, where $f(r) = 0$ has a local minimum. In that case, the Cauchy horizon and event horizon that appear for $m>p$ coincide as $m \rightarrow p$.
The limit is called {\it extremal} since the horizon cloaking the central singularity of the black hole disappears if the charge of the black hole is increased -- the solution therefore represents a limit to the charge a black hole can carry while remaining a ``black hole''. 
Since the metric function $f(r)$ is a quadratic in $1/r$, this naturally corresponds to a repeated root of the function. 
A consequence of this repeated root is that the metric function has a turning point at $f=0$, leading to a zero temperature for the black hole. Once we add the non-linear terms however, the behaviour of the energy momentum tensor is no longer a single power of $r$, and becomes much more complex, meaning whether or not we still have this critical behaviour is not immediately clear. Indeed, if we were to have a genuine extremal limit, with horizons `disappearing' we would most likely be driven to small radius, where the weak-field approximation of our action \eqref{action} would break down (see discussion in Section \ref{WeakFieldSec}).

For the purposes of this subsection (and the ensuing discussions) we will refer to a zero temperature horizon, i.e.\ where $f(r_+) = f'(r_+)=0$, as a {\it local extremal limit}, understanding however that it might not, strictly, represent an actual bound on the charge of the black hole.
The plots of $f(r)$ in Figure \ref{q0f(r)Ex} demonstrate that this type of ``extremal limit'' does indeed occur, but also suggest a peculiar feature; for fixed coupling $a$, there can be cases where, for any mass $m$, no local extremal limit occurs where multiple horizons would coincide. Instead, the corresponding non-linear curves can be monotonic where their linear counterparts become extremal. 
In Appendix \ref{ExtremalJumpDerivations}, we investigate this possibility analytically with the $q = 0$ case. We derive the following necessary and sufficient conditions on the parameters for a local extremal limit to occur for some mass $m$,
\begin{align}
    p^2 \geq \frac{27}{2}a.
\end{align}
Hence, in the non-linear theory, cases with $p^2 < \frac{27}{2}a$ will have no local extremal limit for any $m$.

\begin{figure}[ht]
\centering
\includegraphics[width=0.6\textwidth]{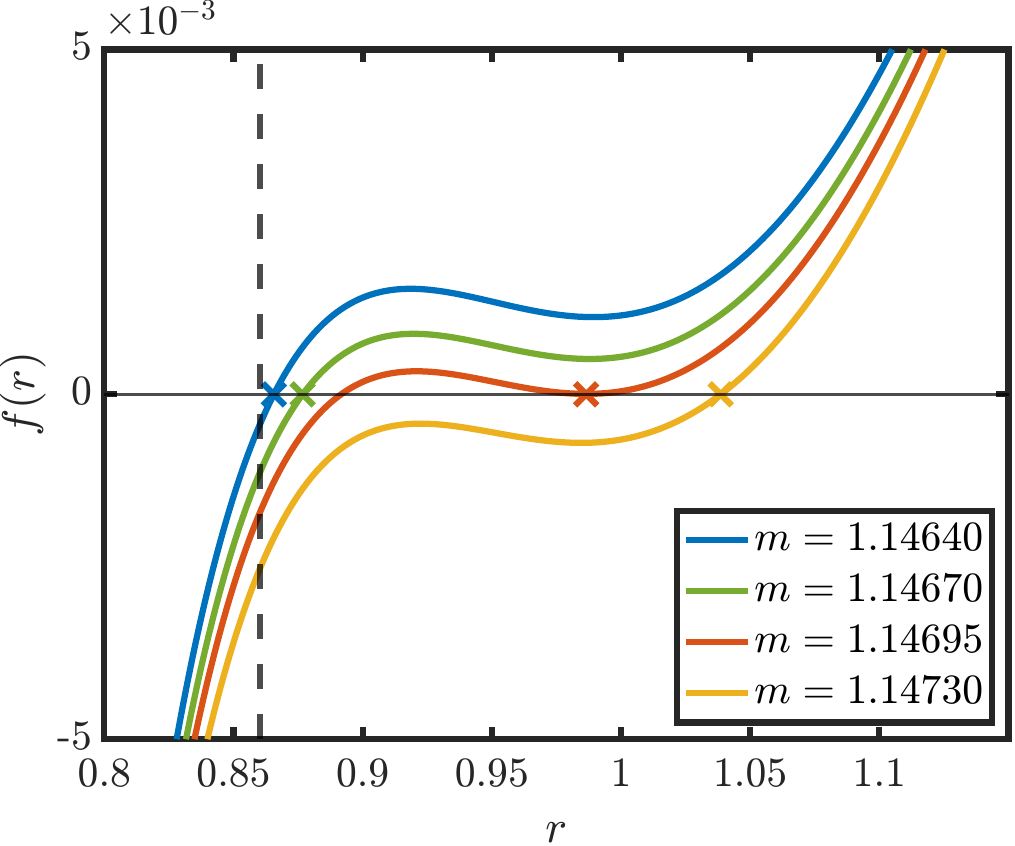}
\caption{A plot exhibiting a jump in black hole event horizon radius $r_+$ as $m$ is varied, for $q = 0$, $p = 1.17$, $a = 0.1$ and $\Lambda = 0$. 
The horizontal thin line marks $f(r) = 0$, where the horizons occur. The vertical dashed line marks the threshold for the DEC violation, so each solution shown conserves DEC outside the horizon.
The black hole horizons in each case are the largest root of $f(r) = 0$, and are denoted by crosses.
Each plot shown has a minimum and as the minimum crosses through $f(r) = 0$, shown for the red curve, there is a sudden jump in the horizon $r_+$. This corresponds to crossing through the local extremal limit.
Note that the green and blue curves also have stationary points \textit{outside} the horizon. 
}
\label{f(r)jump}
\end{figure}

\begin{figure}[ht]
\centering
\includegraphics[width=0.6\textwidth]{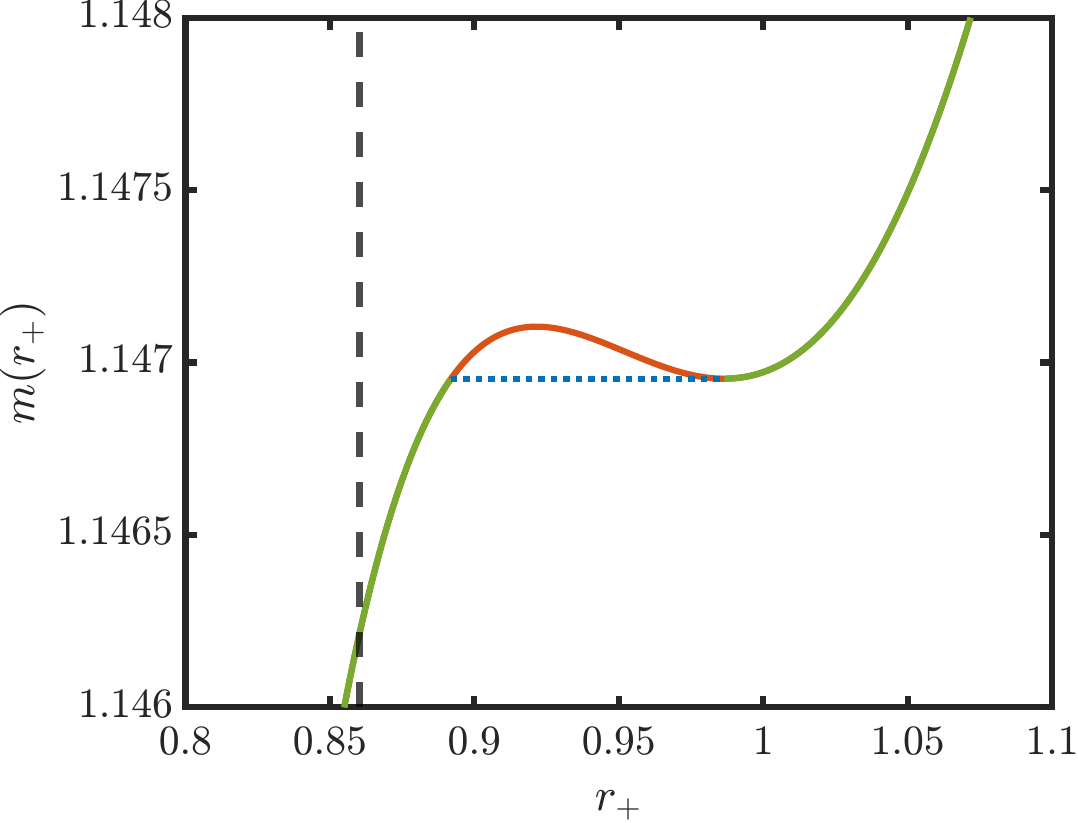}
\caption{A plot of $m(r_+)$, showing how a jump in black hole event horizon radius $r_+$ can occur as $m$ is varied, for $q = 0$, $p = 1.17$, $a = 0.1$ and $\Lambda = 0$. 
The vertical dashed line marks the threshold for the DEC violation, so all $r_+$ above this are valid. 
The green sections of the curve show the allowed horizon radii $r_+$ (modulo DEC). The red section of the curve shows the disallowed horizon radii, since for masses $m$ in this section, there is a larger valid horizon $r_+$.
The horizontal dotted blue line marks a jump between valid event horizon radii.
}
\label{mass_jump}
\end{figure}

In contrast to the linear theory, a lack of a local extremal limit \textbf{does not} imply that there are no horizons at all. Instead, when becoming locally super-extremal (say by changing the mass $m$), other locations where $f(r) = 0$ could become the event horizon, provided that the DEC is not violated. An example is shown in Figure \ref{f(r)jump}.
However, in this scenario (as we see in Figure \ref{f(r)jump}) it appears that the physical event horizon $r_+$ ``jumps'' as $m$ is varied, offering another peculiar feature of the non-linear theory. 
Note, this is not a dynamical jump, rather, a discontinuity in the locus of the horizon as we increase the mass incrementally.
To investigate this, we express the mass $m$ as a function of horizon radius $r_+$,
\begin{align}
    m(r_+) = \frac{r_+}{2} + \frac{p^2}{2r_+} - \frac{ap^4}{5r_+^5}.
\end{align}
An example of $m(r_+)$ with a jump in $r_+$ as $m$ is varied is shown in Figure \ref{mass_jump}.
We see that in order to have a jump in horizon radius, the function $m(r_+)$ needs to be non-monotonic. Furthermore, the DEC must be satisfied during this non-monotonic region. Using these criteria, in Appendix \ref{ExtremalJumpDerivations}, we derive the range of parameters for which a jump in horizon radius can occur. The resulting range of $p^2$ for a jump is,
\begin{align}
   \frac{27}{2}a < p^2 \leq \eta \, a,
\end{align}
where $\eta \approx 13.987$.
Note the case shown in Figures \ref{f(r)jump} and \ref{mass_jump} satisfies this requirement.

We summarise the results of this section as follows:
\begin{itemize}
\item[1)]
If $p^2 < \frac{27}{2}a$, then $f(r)$ is monotonic for all $m$, and there is no jump, nor an extremal limit (except for $p^2 = 27a/2$, where \textit{three} horizons coincide), with the DEC providing the only lower bound on $m$.
\item[2)]
If $\frac{27}{2}a < p^2 \leq \eta \, a$, for some $m$ there is a local extremal limit where two horizons coincide, but the horizon $r_+$ is discontinuous as $m$ is varied, while conserving DEC.
\item[3)]
If $p^2 > \eta \, a$, there is an extremal limit for some $m$ in the ordinary sense. The horizon $r_+$ is continuous as $m$ is varied, to prevent DEC violation. 
\end{itemize}

Another curious feature is that some of the curves in Figure \ref{f(r)jump} also have stationary points \textit{outside} the horizon. This possibility has been noted previously in another NLED theory \cite{Liu:2019rib}. 
This feature only occurs here in a narrow range of masses before DEC is violated. 

Note that the strong energy condition $r_+^4 \leq 6ap^2$ would rule out any of the interesting features observed in this section. However, as previously discussed, we take the view that the SEC is an excessive physical condition, and hence we allow its violation.

\section{Thermodynamics} \label{ThermoOverallSec}

Since the conception of Hawking radiation \cite{hawking1975}, the interpretation of black holes as genuine thermodynamic systems has evolved from mathematical analogy to compelling physical reality. 
This paradigm has been significantly advanced through the study of black hole phase transitions, which exhibit striking parallels with conventional thermodynamic systems. The analysis of these transitions fundamentally relies on the Gibbs free energy $G = M - TS$, where $M$ represents the enthalpy within the extended phase space formalism \cite{kastor2009, mann2017}.

As is well established in the literature \cite{chamblin1999, Kubiznak:2012wp}, the Reissner-Nordström anti-de Sitter (RNAdS) solution exhibits characteristic swallowtail behaviour in phase plots, displaying both first and second-order phase transitions. In this section, we demonstrate how NLED modifications give rise to richer phase transition phenomena, including additional turning points, and higher-order catastrophes.

\subsection{A primer on phase transitions} \label{CritRNAdSSec}

Before we investigate the phase structure of the NLED system, for reference we first briefly remind the reader of the phase properties for RNAdS black holes within the linear Maxwell theory.

For a given temperature $T$, the thermodynamically preferred state of the system is the one with lowest Gibbs free energy $G$, provided this state is stable. Hence, it is possible to have \textit{phase transitions} between states if the curve $G(T)$ crosses itself. We therefore can understand the thermodynamic phase structure of black holes by plotting phase diagrams, where the Gibbs free energy $G$ is plotted as a function of temperature $T$. We will discuss stability in the context of heat capacity in Section \ref{heat_cap_sec}.

For RNAdS, the relevant quantities for phase plots are,
\begin{align} \label{LinearTemp}
    T&=\frac{1}{4\pi r_+}\left(1-\Lambda r_+^2-\frac{q^2+p^2}{r_+^2}\right),\\ 
    G&=\frac{r_+}{4}+\frac{r_+^3\Lambda}{12}+\frac{3}{4}\frac{q^2+p^2}{r_+}.
\end{align}
We give more details on how to obtain these quantities when we determine them for the non-linear theory in Section \ref{NonLinThermoSec}. Note that $T$ and $G$ exhibit the electromagnetic duality $p \leftrightarrow q$ in RNAdS.

We can use the horizon radius $r_+$ as a parameter for plotting the resulting $(G, T)$ curve, however there are some restrictions on $r_+$ to ensure that it corresponds to a black hole event horizon. The two conditions are: (i) the corresponding mass $m$ must be positive, and (ii) $r_+$ must be the largest valid horizon with positive temperature (for example it cannot correspond to an inner horizon).

For RNAdS, rearranging the horizon condition for the mass gives,
\begin{align}
    m = \frac{r_+}{2} \left(1 - \frac{\Lambda}{3} r_+^2 + \frac{q^2+p^2}{r_+^2}\right).
\end{align}
Provided $\Lambda < 0$ and $q^2+p^2 > 0$, it is immediate that $m > 0$ for all $r_+ > 0$, so (i) is guaranteed in this case. To check (ii), note $m$ has the following extreme behaviour,
\begin{equation}
   m \sim
   \begin{cases}
    (q^2+p^2)r_+^{-1}/2 \, , \quad &\text{for small } r_+,\\
    (-\Lambda)r_+^{3}/3 \, ,\quad &\text{for large } r_+.
\end{cases}
\end{equation}
Between these extremes, $m(r_+)$ has a single stationary point at,
\begin{align} \label{ConditioniiRNAdS}
    r_+^2 = \frac{-1 + \sqrt{1+4(-\Lambda)(q^2+p^2)}}{2(-\Lambda)},
\end{align}
which corresponds to the minimum mass to prevent a naked singularity.
In this simple case, this stationary point provides the minimum valid event horizon radius. When we add in the non-linear electromagnetism, the event horizon validity conditions will generally be determined numerically.
In any analysis we conduct from here on, we ensure that any turning points we report on correspond to a valid event horizon, and hence satisfy conditions (i) and (ii). In the non-linear theory, we will also have to ensure the DEC is satisfied, as discussed in Section \ref{DECVioSec}.

\begin{figure}[ht]
\centering
\includegraphics[width=0.6\textwidth]{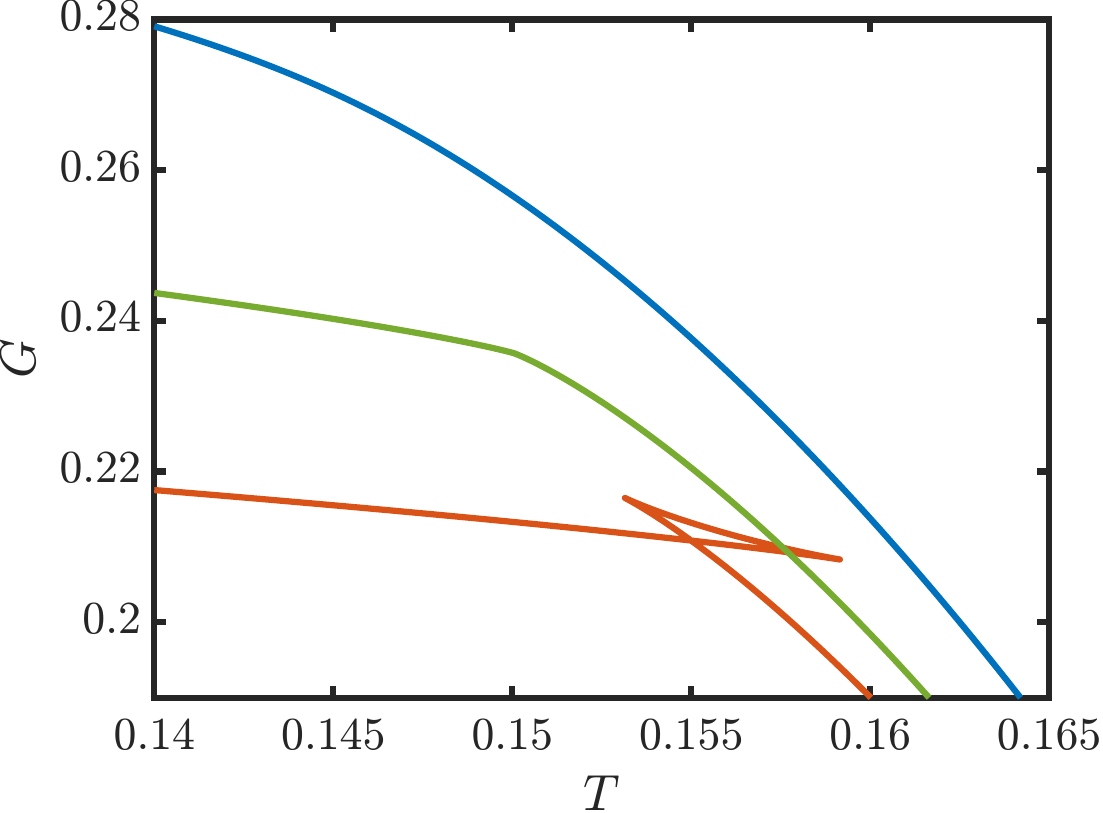}
\caption{A plot showing phase diagrams of Gibbs free energy $G$ against temperature $T$, for three cases in the linear theory with RNAdS with $\Lambda = -1$. Due to the symmetry in $(p,q)$ in the linear theory, we have set $p = 0$ in these diagrams. The different cases correspond to no turning points ($q = 0.35$, blue), one turning point ($q = 1/\sqrt{12}$, green) and two turning points ($q = 0.25$, red). The curves continue in a similar fashion for larger $T$, and for smaller $T$ until $T = 0$ is reached.}
\label{RNAdSPhaseEx}
\end{figure}

Since the thermodynamically-preferred branch of the system is the one with minimal Gibbs free energy (for a given temperature), phase transitions can occur whenever the $G(T)$ curve intersects itself, as $r_+$ is varied.
An example of a phase transition in RNAdS is shown in the red curve of Figure \ref{RNAdSPhaseEx}, demonstrating also a `swallowtail' shape. This shape inherits its name from the projection of shapes within catastrophe theory \cite{chamblin1999, poston1996catastrophe}.

The curve of $G(T)$ is allowed to cross itself due to the presence of \textit{turning points}, where the curve splits into different branches as the temperature is varied.
The location $r_+$ of turning points satisfy $dT/dr_+ = 0$, or equivalently $dG/dr_+ = 0$, giving an effective quadratic for the RNAdS case,
\begin{align}
    (-\Lambda) r_+^4 - r_+^2 + 3(q^2+p^2) = 0.
\end{align}
Hence, there are at most two possible solutions,
\begin{align}
    r_+^2 = \frac{1 \pm \sqrt{1-12(-\Lambda)(q^2+p^2)}}{2(-\Lambda)},
\end{align}
depending on the sign of $1-12(-\Lambda)(q^2+p^2)$. It can be checked that for $\Lambda < 0$ and $q^2+p^2 > 0$, these turning points are always beyond the threshold set by condition (ii) in \eqref{ConditioniiRNAdS}, and are therefore valid event horizon radii. Examples of phase diagrams with 0, 1 and 2 turning points for RNAdS are shown in Figure \ref{RNAdSPhaseEx}.

When phase transitions occur, they have an associated `order', as in ordinary thermodynamic systems. For our purposes, the order of a phase transition is defined to be the lowest-order derivative of $G(T)$ that is discontinuous at the phase transition. For example, in Figure \ref{RNAdSPhaseEx}, the red curve shows a first-order phase transition where the two branches intersect, the green curve shows a second-order phase transition, and the blue curve shows a single phase devoid of transitions. 

We can probe the phase structure of black holes in NLED theories by asking about the number of possible turning points. Scenarios with more turning points than RNAdS are likely to have a richer phase transition structure. Indeed, NLED can give rise to additional turning points, and we investigate this in subsequent sections, after first deriving the thermodynamics.

\subsection{Thermodynamic results in non-linear theory} \label{NonLinThermoSec}

With the introduction of the non-linear terms in the action \eqref{action}, the thermodynamics will receive appropriate modifications. In particular, the first law and Smarr relation gain terms due to the new Lagrangian couplings $a,b$. This has been discussed in \cite{Bokulic:2021dtz}, where the rationale for including these additional terms is explored. 

In our analysis, we will allow variations $da$, $db$ in the first law, and permit the possibility that there is some \textit{as yet unknown} physical mechanism where these variations are non-vanishing. For example, in extended thermodynamics \cite{mann2017}, the cosmological constant $\Lambda$ can vary as the potential of a scalar field.
If we are more restrictive, and force $a,b$ to be fixed, we can take $da, db = 0$, and the remainder of the first law will still apply.

In light of this, we can calculate the first law by varying $f(r_+) = 0$ in \cref{frfull} with respect to $(r_+, M, \Lambda, p, q, a, b)$,
\begin{equation} \label{FullFirstLaw}
    dM = T dS + \Phi_e dq + \Phi_m dp - \chi_a da - \chi_b db + V dP,
\end{equation}
where the thermodynamic quantities appearing are,
\begin{equation} \label{GeneralThermoVariables}
\begin{aligned}
M& = m, \quad S= \pi r_{+}^2, \quad P=-\frac{\Lambda}{8\pi},\quad V=\frac{4}{3}\pi r^3_+, \\
    T & = \frac{f'(r_+)}{4\pi} = \frac{1}{4\pi} \left[\frac{1}{r_+} -\Lambda r_+ + \frac{1}{2} \left(r_+ + \frac{4p^2}{r^3_+}(2b-a) \right) F_{rt}^2 - \frac{3q}{2r_+} F_{rt} + \frac{2a p^4}{r^7_+} - \frac{p^2}{r^3_+}\right], \\
    \Phi_e & = \frac{1}{4} \int^{r_+} dr \left[\left(\left(2r^2 + \frac{8p^2}{r^2}(2b-a)\right)F_{rt} - 3q \right)  \frac{\partial F_{rt}}{\partial q} - 3 F_{r t} \right], \\
    \Phi_m & = \frac{p}{r_+} -\frac{4ap^3}{5r_+^5} + \frac{1}{4} \int^{r_+} dr \left[\left(\left(2r^2 + \frac{8p^2}{r^2}(2b-a)\right)F_{rt} - 3q \right)  \frac{\partial F_{rt}}{\partial p} + \frac{8p}{r^2}(2b-a) F_{rt}^2 \right], \\
    \chi_a & = \frac{p^4}{5 r_+^5} - \frac{1}{4} \int^{r_+} dr \left[\left(\left(2r^2 + \frac{8p^2}{r^2}(2b-a)\right)F_{rt} - 3q \right)  \frac{\partial F_{rt}}{\partial a} - \frac{4p^2}{r^2} F_{rt}^2 \right],\\
    \chi_b & = - \frac{1}{4} \int^{r_+} dr \left[\left(\left(2r^2 + \frac{8p^2}{r^2}(2b-a)\right)F_{rt} - 3q \right)  \frac{\partial F_{rt}}{\partial b} + \frac{8p^2}{r^2} F_{rt}^2 \right].
\end{aligned}
\end{equation}

The Smarr relation is,
\begin{align} \label{SmarrGeneral}
    m&=2(TS-PV-a\chi_a-b\chi_b)+q\Phi_e+p\Phi_m,
\end{align}
where the new terms proportional to $a,b$ appear with the prefactor $2$, due to Euler's theorem for homogeneous functions \cite{mann2017, Hu:2018njr}, since $a, b \sim \text{[length]}^2$.

When we come to analyse phase diagrams, we will also need the Gibbs free energy, defined via Legendre transformation,
\begin{equation}
\begin{aligned}
    G = M & - TS\\ 
    = \frac{r_+}{4} & + \frac{r_+^3 \Lambda}{12} + \frac{3}{4} \frac{p^2}{r_+} - \frac{7}{10}\frac{a p^4}{r_+^5} - \frac{1}{8}\left(r_+^3 + \frac{4(2b-a)p^2}{r_+}\right)F_{rt}^2 + \frac{3qr_+}{8} F_{rt} \\ 
    & + \frac{1}{4} \int^{r_+} \left[\left(r^2 + \frac{4(2b-a)p^2}{r^2}\right)F_{rt}^2 - 3q F_{rt}\right] dr  \; .
\end{aligned}
\end{equation}

In the following sections, we will investigate the consequences of the non-linear terms on phase diagrams, and the number of possible turning points, through analytic and numerical approaches.

\subsection{Perturbative analysis} \label{PertSec}

We have framed our general thermodynamic results thus far in terms of $F_{rt}$, which in principle can be obtained from the cubic \eqref{Fcubic}. A deeper understanding of $F_{rt}$ then enhances our appreciation of the black hole solution. When $a = b = 0$, \eqref{Fcubic} is solved by the linear Maxwell solution $F_{rt} = q/r^2$, as expected. Since the action \eqref{action} is only a weak-field correction, where in principle the couplings should be small, we can find an approximate solution for $F_{rt}$ by considering a perturbative expansion in $a,b$ of the solution to the cubic \eqref{Fcubic}.

To linear order in $a$ and $b$, the physical solution to the cubic \eqref{Fcubic} is,
\begin{align} \label{PertFrtSol}
    F_{rt}&=-F_{tr}=\frac{q}{r^2}\left(1-4\frac{aq^2+(2b-a)p^2}{r^4}\right).
\end{align}

However, since this is a series expansion of the full solution, this expansion can fail at small $r$, no matter how small $a$ and $b$ are, as the correction term can dominate over the background term $q/r^2$. Hence, this solution will only generally be valid for sufficiently large $r$. Although, if this breakdown occurs inside the resulting black hole event horizon $r_+$, then this will not affect our analysis.

We can use the next term in the expansion series as a measure of the error of truncating at a fixed term in the expansion. If we intend for the approximation to be valid with a relative error of at most $\epsilon$, then the approximation breaks down at roughly,
\begin{equation} \label{PertBreakdown}
    \begin{aligned}
    r \lesssim \; \left[\frac{4(aq^2+(2b-a)p^2)}{\epsilon}\right]^{1/4} \quad &\text{(linear Maxwell solution)}, \\
    r \lesssim \; \left[\frac{16(aq^2+(2b-a)p^2)(3aq^2+(2b-a)p^2)}{\epsilon} \right]^{1/8} \quad &\text{(perturbative solution)}.
\end{aligned}
\end{equation}
We could choose, for example, $\epsilon = 0.1$ to provide sufficient accuracy. This allows us to take $a,b$ sufficiently small, so that the perturbative solution \eqref{PertFrtSol} is still accurate everywhere outside the event horizon, but not so small that it gives an insignificant improvement over the linear Maxwell solution.
In fact, there is a range of parameter space that can allow for an increased maximum number of turning points, compared to the linear Maxwell case.

With the perturbative form \eqref{PertFrtSol} for $F_{rt}$, we can start to identify simple explicit expressions for the black hole solution. To linear order in $a$ and $b$, the metric function $f(r)$ is,
\begin{align}
    f(r)&=1-\frac{2m}{r}-\frac{\Lambda r^2}{3}+\frac{q^2+p^2}{r^2}-\frac{2}{5}\frac{aq^4+2(2b-a)p^2q^2+ap^4}{r^6}\label{f(r)solution},
\end{align}
and the thermodynamic variables \eqref{GeneralThermoVariables} reduce to,
\begin{equation}
\begin{aligned} \label{PertThermoVariables}
    M& = m, \quad S= \pi r_{+}^2, \quad P=-\frac{\Lambda}{8\pi},\quad V=\frac{4}{3}\pi r^3_+, \\
    T&=\frac{f'(r_+)}{4\pi}=\frac{1}{4\pi r_+}\left(1-\Lambda r_+^2-\frac{q^2+p^2}{r_+^2}+2\frac{a(p^2-q^2)^2 + 4b p^2 q^2}{r_+^6}\right),\\ 
    \Phi_e &=\frac{q}{r_+}\left(1-\frac{4}{5}\frac{aq^2+(2b-a)p^2}{r_+^4}\right), \quad \Phi_m=\frac{p}{r_+}\left(1-\frac{4}{5}\frac{ap^2+(2b-a)q^2}{r^4_+}\right),\\
     \chi_a &=\frac{(q-p)^2 (q+p)^2}{5r_+^5},\quad \chi_b=\frac{4 q^2 p^2}{5r_+^5}.
\end{aligned}
\end{equation}

Lastly the Gibbs free energy is,
\begin{align}
    G=\frac{r_+}{4}+\frac{r_+^3\Lambda}{12}+\frac{3}{4}\frac{q^2+p^2}{r_+}-\frac{7}{10}\frac{a\left(p^2-q^2\right)^2+4bp^2q^2}{r_+^5}.
\end{align}

Note that at first perturbative order, the expression for $f(r)$, and many of the thermodynamic quantities exhibit the electromagnetic duality $p \leftrightarrow q$, just as in vacuum linear electromagnetism.

\begin{figure}[t]
    \centering
        
    \begin{minipage}{0.47\textwidth}
        \includegraphics[width=\textwidth]{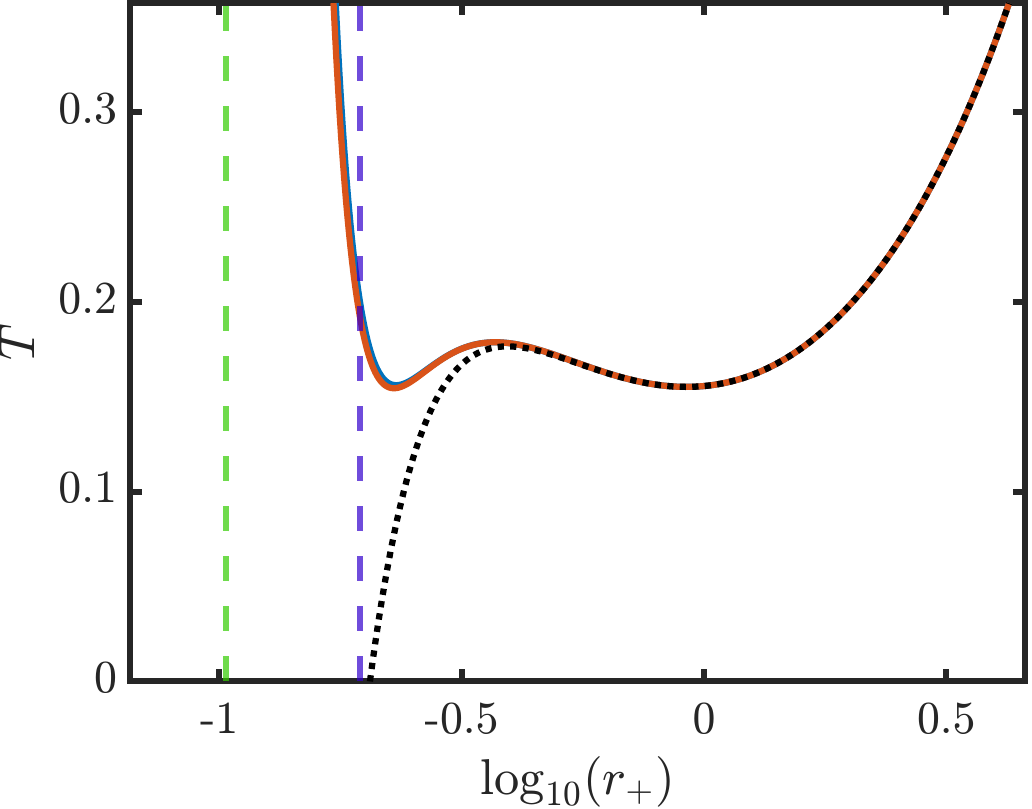}
    \end{minipage}
    \hspace{0.4cm}
    \begin{minipage}{0.47\textwidth}
        \includegraphics[width=\textwidth]{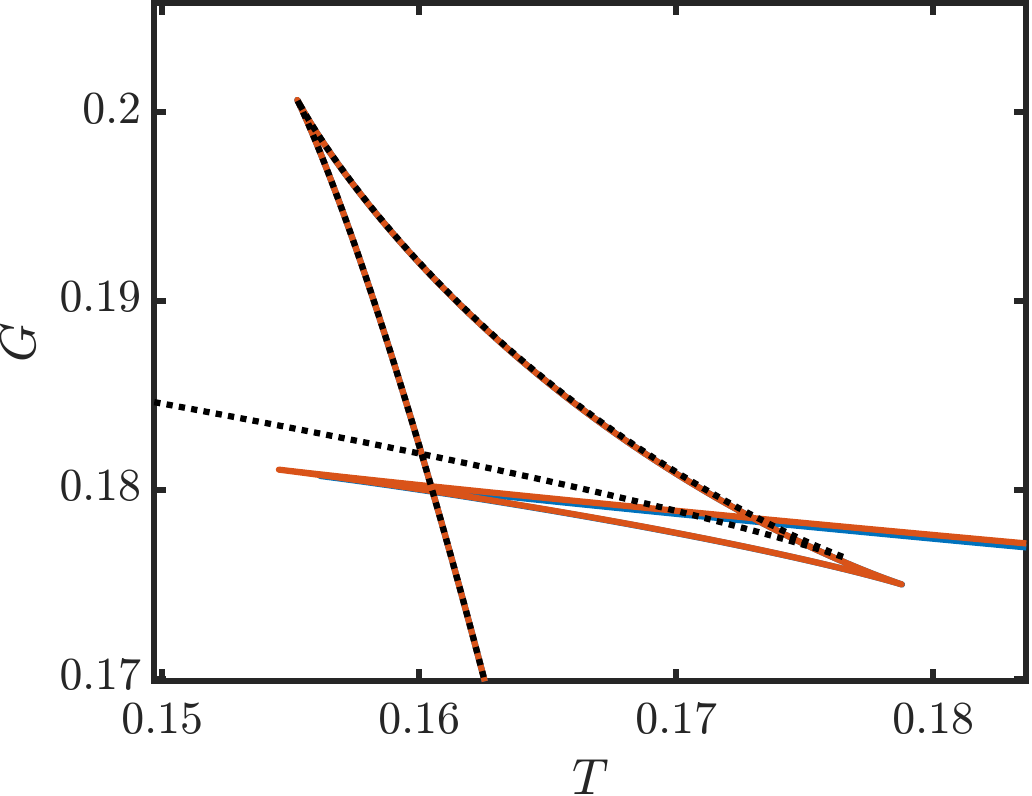}
    \end{minipage}
    
    \caption{Plots demonstrating a case with three turning points. Left: a plot of temperature $T$ with horizon radius $r_+$ for $q = 0.06$, $p = 0.2$, $a = 0.0094$, $b = 0.007$, $\Lambda = -1$. The green dashed line is the event horizon threshold, and the purple dashed line is the DEC violation threshold, which occurs just to the left of the smallest $r_+$ turning point. Right: a phase diagram of Gibbs free energy $G$ against temperature $T$ for the same choice of parameters. Both: The red curves are the numerical results, the blue are the perturbative results (which almost overlap, showing the validity of the perturbative analysis for this set of parameters), and the black dotted curves are the analogous results for the linear RNAdS regime. Note that since $\Lambda = -1$, $r_+$ is dimensionless.}
    \label{LinearisedPlotsEx}
\end{figure}

For large $r$, $F_{rt} \approx q/r^2$, and $T$ and $G$ also approach their RNAdS expressions, provided $r$ is within the regime of validity \eqref{PertBreakdown}. This means that any additional turning points compared to RNAdS (as discussed in Section \ref{CritRNAdSSec}) must come at sufficiently small $r$ that the linear Maxwell solution is inaccurate, but not too small that the perturbative approximation also breaks down. Later we will do a full numerical analysis, which allows us to numerically check the validity of our results obtained perturbatively here. Within this regime, it is possible to generate cases with three turning points accurately from this analysis, with any breakdown appearing at a lower $r_+$ than the event horizon threshold.
An example is shown in Figure \ref{LinearisedPlotsEx}, where we see three turning points, one more than the maximum possible in the linear theory. The DEC violation boundary appears to be close to the smallest $r_+$ turning point, and below that DEC boundary, the results might be untrustworthy, perhaps requiring higher order terms in the action \eqref{action}. However we will consider the turning points in Figure \ref{LinearisedPlotsEx} to be valid, since they occur within the DEC `safe' region.
In the language of catastrophe theory, the shape in Figure \ref{LinearisedPlotsEx} would be considered a butterfly catastrophe, a higher-order catastrophe than the swallowtail observed in RNAdS thermodynamics \cite{chamblin1999, poston1996catastrophe}.

\subsection{$a = 0$ case} \label{a0Sec}

The perturbative analysis in Section \ref{PertSec} is restricted to sufficiently small $a,b$. There may be features of the solutions that are not elucidated by this restriction. 
Having a simple solution for $F_{rt}$ allows us to make further analytic progress with determining the other properties of the black hole solution, and its thermodynamics.
In Section \ref{Analyticq0Sec}, we discussed a simple analytic solution obtained by setting the electric charge $q = 0$.
In this section, we consider a different tractable case, where NLED corrections are important, but instead the dyonic feature of the black holes is maintained. Here, we set $a = 0$ in the action \eqref{action}, dropping the $F^2$ term. This may not physically relate to a particular NLED theory, but mathematically it allows us to probe the rich thermodynamic structure of NLED black holes analytically.

When $a = 0$, the cubic \eqref{Fcubic} for $F_{rt}$ reduces to a linear equation, giving the \textit{exact} solution,
\begin{equation} \label{Frta0eq}
    F_{rt} = -F_{tr} = \frac{q r^2}{r^4 + 8bp^2},
\end{equation} 
where $b \geq 0$ is unrestricted in magnitude, in contrast to the solution in Section \ref{PertSec}.
In fact, there is no DEC violation in this case, as seen explicitly from equations \eqref{rho_positive} and \eqref{rho_plus_p_positive}.

\begin{figure}[ht]
\centering
\includegraphics[width=0.5\textwidth]{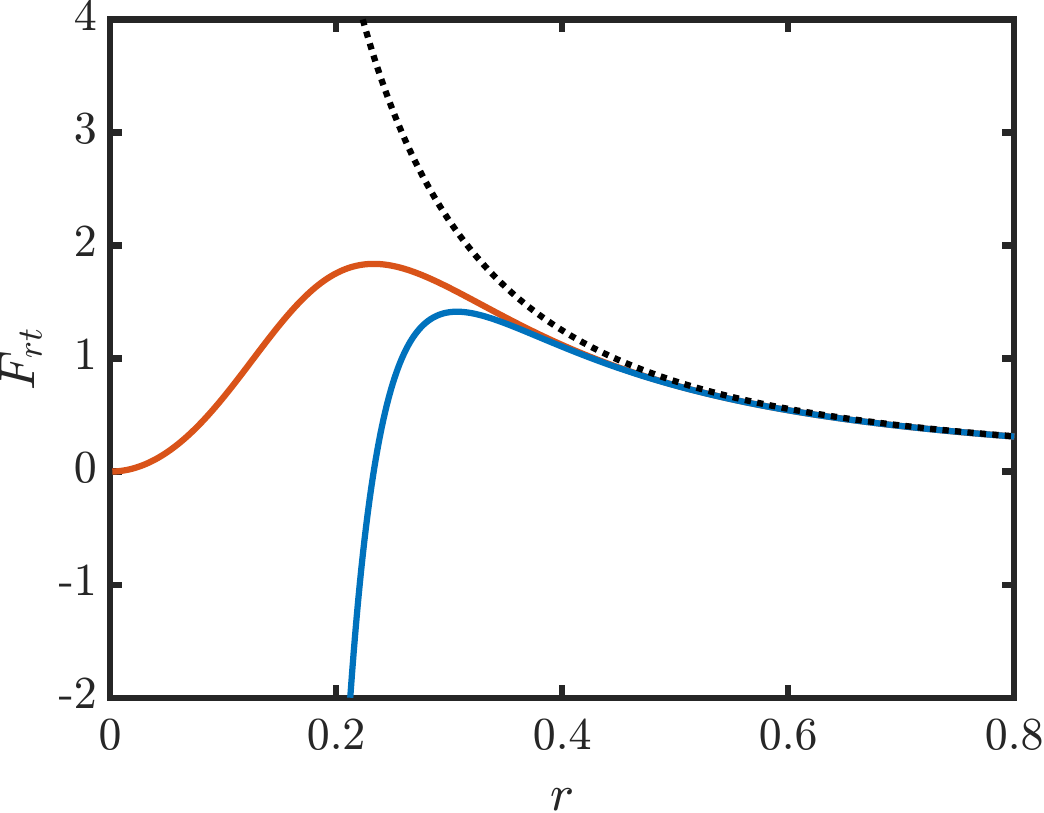}
\caption{A plot of $F_{rt}$ against $r$ for $q = 0.2$, $p = 0.1$, $a = 0$ and $b = 0.037$. The red curve is the exact result in \eqref{Frta0eq}, the blue is the perturbative result, and the black dotted curve is the analogous results for the linear (Maxwell) regime. There is no DEC violation for $r > 0$ since $a = 0$.}
\label{a0FrtEx}
\end{figure}

Note that in this solution \eqref{Frta0eq}, $F_{rt} \sim r^2$ as $r \rightarrow 0$, and hence is \textit{regular} there. This is in sharp contrast to the linear regime, and the perturbative solution analysed in Section \ref{PertSec}. A example plot of this solution is shown in Figure \ref{a0FrtEx}. We can see that for large $r$, the linear and the perturbative results are good approximations for the true result, but that for sufficiently small $r$, they both fail. The true numerical result is regular as $r \rightarrow 0$, as opposed to divergent with the linear and perturbative results.

For this analytically tractable case, the metric function $f(r)$ can be found as,
\begin{align} \label{a0fmetric}
    f(r)=1&-\frac{2m}{r}-\frac{\Lambda r^2}{3}+\frac{p^2}{r^2} - \frac{1}{r} \Xi(r),
\end{align}
where,
\begin{equation}
\begin{aligned}
    \Xi(r) \equiv & \int^r dr' \frac{q^2 r'^2}{r'^4 + 8bp^2} \\
    = & \frac{q^2}{8(2bp^2)^{1/4}}\left[2\left(\tan^{-1}\left(\frac{\sqrt{2}r}{(8bp^2)^{1/4}}+1\right)\right. +\tan^{-1}\left(\frac{\sqrt{2}r}{(8bp^2)^{1/4}}-1\right)\right)   \\
    &\quad \quad \quad \quad \quad  \quad \; \;  \; +\left.\ln\left(\frac{r^2-2(2bp^2)^{1/4}\, r+(8bp^2)^{1/2}}{r^2+2(2bp^2)^{1/4}\, r+(8bp^2)^{1/2}}\right)-2\pi\right],
\end{aligned}
\end{equation}
assuming $bp^2 \neq 0$. Nonetheless, in the $b \rightarrow 0$ limit, this expression recovers RNAdS. The final $2\pi$ arises from the integral, to ensure no contribution to the black hole mass in $f(r)$ as $r \rightarrow \infty$.

The thermodynamic variables \eqref{GeneralThermoVariables} reduce to,
\begin{equation}
\begin{aligned} \label{a0thermovariables}
    M = & \, m, \quad S= \pi r_{+}^2, \quad P=-\frac{\Lambda}{8\pi},\quad V=\frac{4}{3}\pi r^3_+, \\
    T= &\, \frac{f'(r_+)}{4\pi}=\frac{1}{4\pi r_+}\left(1-\Lambda r_+^2-\frac{p^2}{r_+^2} - \frac{q^2 r_+^2}{r_+^4 + 8bp^2}\right), \\ 
    \Phi_e = & -\frac{\Xi(r_+)}{q}, \quad \Phi_m = \frac{p}{r_+}\left(1 + \frac{1}{4p^2} \frac{q^2 \, r_+^4}{r_+^4 + 8bp^2} + \frac{r_+ \Xi(r_+)}{4p^2}\right), 
     \\
    \chi_b = & - \frac{1}{8 b} \left( \frac{q^2 \, r_+^3}{r_+^4 + 8bp^2} +  \Xi(r_+)\right). 
\end{aligned}
\end{equation}
Restricting ourselves to the family of solutions with $a = 0$ would give a first law that does not feature a $da$ term, and hence the $\chi_a$ has been omitted above.

In this case, we see an explicit breaking of the electromagnetic duality $p \leftrightarrow q$ in the expression for $f(r)$, and the thermodynamic quantities that are usually symmetric in vacuum linear electromagnetism. As such, the turning point analysis will generally produce different results under $p \leftrightarrow q$. 

The Gibbs free energy is,
\begin{align}
G=\frac{r_+}{4}+\frac{r_+^3\Lambda}{12}+\frac{3}{4}\frac{p^2}{r_+} + \frac{q^2 r_+^3}{4(r_+^4 + 8bp^2)} - \frac{\Xi(r_+)}{2} .
\end{align}

\begin{figure}[t]
    \centering
        
    \begin{minipage}{0.47\textwidth}
        \includegraphics[width=\textwidth]{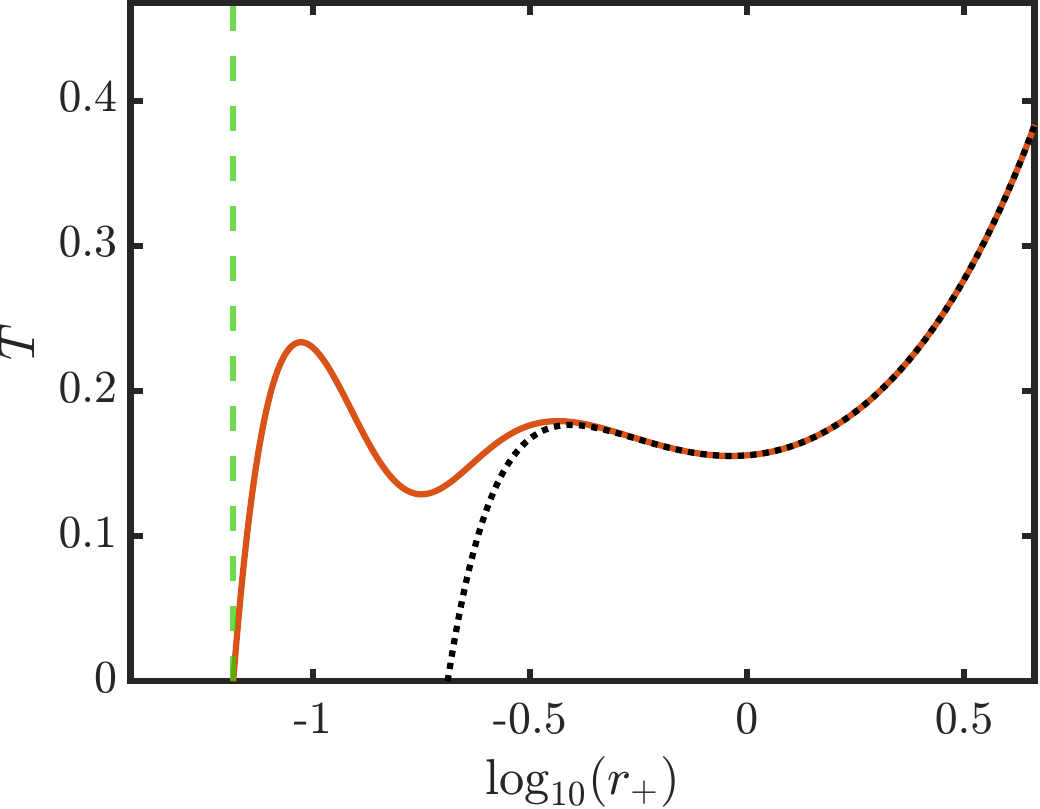}
    \end{minipage}
    \hspace{0.4cm}
    \begin{minipage}{0.47\textwidth}
        \includegraphics[width=\textwidth]{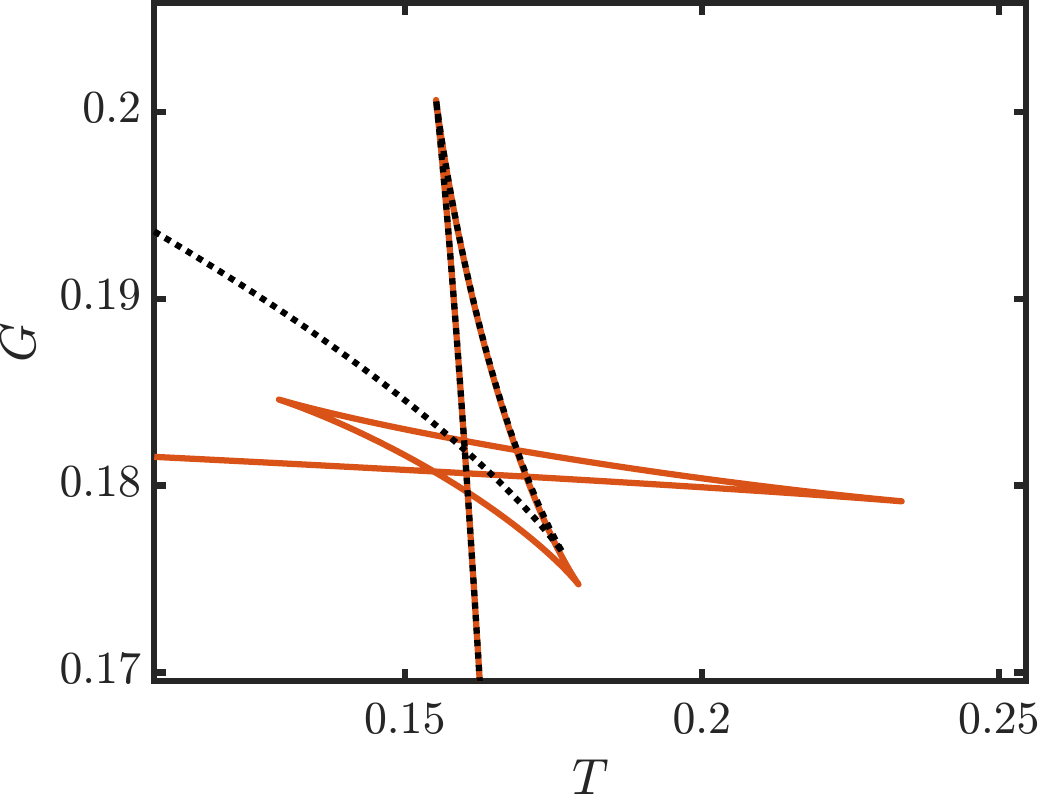}
    \end{minipage}
    
    \caption{Plots demonstrating a case with four turning points. Left: a plot of temperature $T$ with horizon radius $r_+$ for $q = 0.2$, $p = 0.06$, $a = 0$, $b = 0.035$, $\Lambda = -1$. The green dashed line is the event horizon threshold, and there is no DEC violation for $r_+ > 0$ since $a = 0$. Right: a phase diagram of Gibbs free energy $G$ against temperature $T$ for the same choice of parameters. Both: The red curves are the numerical results and the black dotted curves are the analogous results for the linear RNAdS regime. Note that since $\Lambda = -1$, $r_+$ is dimensionless.
    }
    \label{a0PlotsEx}
\end{figure}

In the $a = 0$ case, any turning points that exist satisfy the equation $\frac{dT}{dr_+} = 0$, reducing to,
\begin{equation}
\begin{aligned}
    & (-\Lambda)r_+^{12} - r_+^{10} + (3p^2+3q^2+16(-\Lambda)bp^2)r_+^8-16bp^2 r_+^6 \\
    & + (48bp^4-8bp^2q^2+64(-\Lambda)b^2p^4)r_+^4 - 64b^2p^4r_+^2 + 192b^2p^6 = 0, 
\end{aligned}
\end{equation}
which is a sixth degree polynomial in $r_+^2$, and hence there are at most six turning points in this case.
In practice, these roots would need to satisfy $r_+^2 \geq 0$ to be real, and the resulting $r_+$ must exceed the threshold for a valid event horizon.
An extensive numerical search has revealed up to four turning points in this case.
An example with four turning points is shown in Figure \ref{a0PlotsEx}. In catastrophe theory, this shape would be considered even higher-order than the butterfly observed in Figure \ref{LinearisedPlotsEx}.

\subsection{Numerical analysis}\label{numan}

Our existing analyses of turning points in Sections \ref{PertSec} and \ref{a0Sec} offer analytic control, but they have limitations in their range of validity.
In this section, we numerically solve the cubic \eqref{Fcubic} for $F_{rt}$, and present some results of additional turning points that cannot be captured by our previous analytic results.

For example if $a \neq 0$, then our analysis in Section \ref{a0Sec} does not apply, and we would otherwise be left with the perturbative analysis of Section \ref{PertSec}.
However, provided we are in the non-linear regime ($a \neq 0$ or $b \neq 0$) and the electric charge is non-vanishing ($q \neq 0$), the perturbative solution \eqref{PertFrtSol} for $F_{rt}$ is guaranteed to fail at some small $r$.

\begin{figure}[t]
    \centering
        
    \begin{minipage}{0.47\textwidth}
        \includegraphics[width=\textwidth]{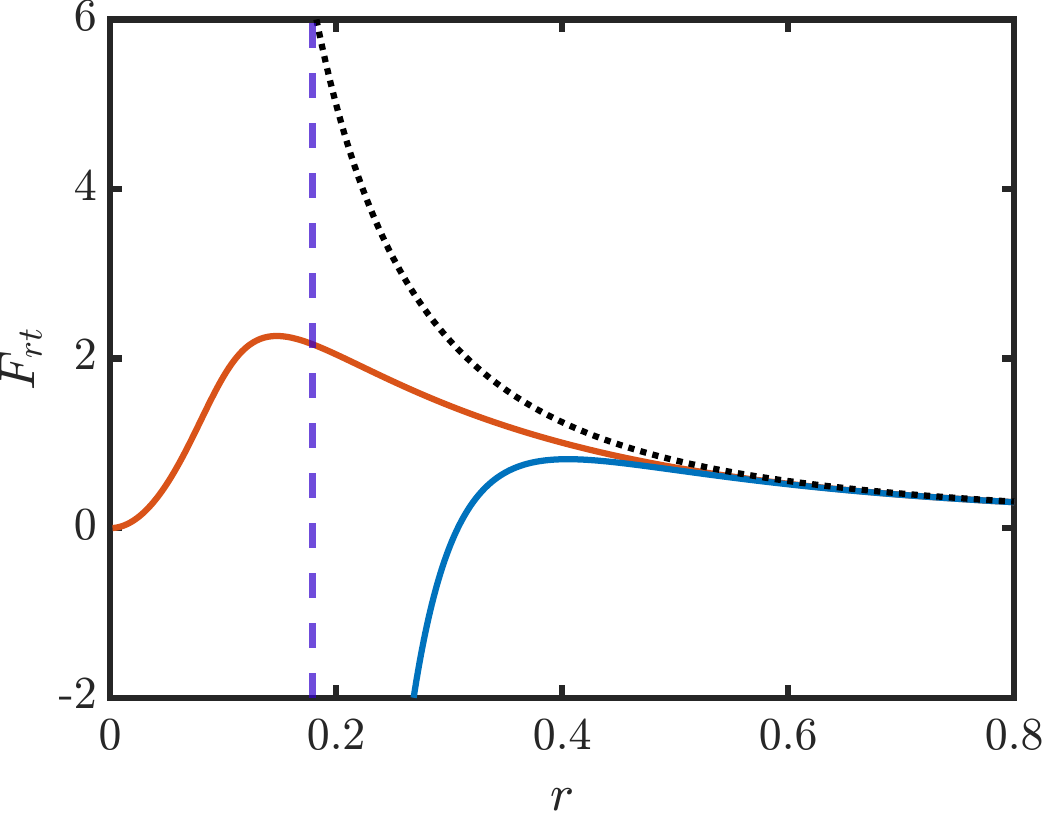}
    \end{minipage}
    \hspace{0.4cm}
    \begin{minipage}{0.47\textwidth}
        \includegraphics[width=\textwidth]{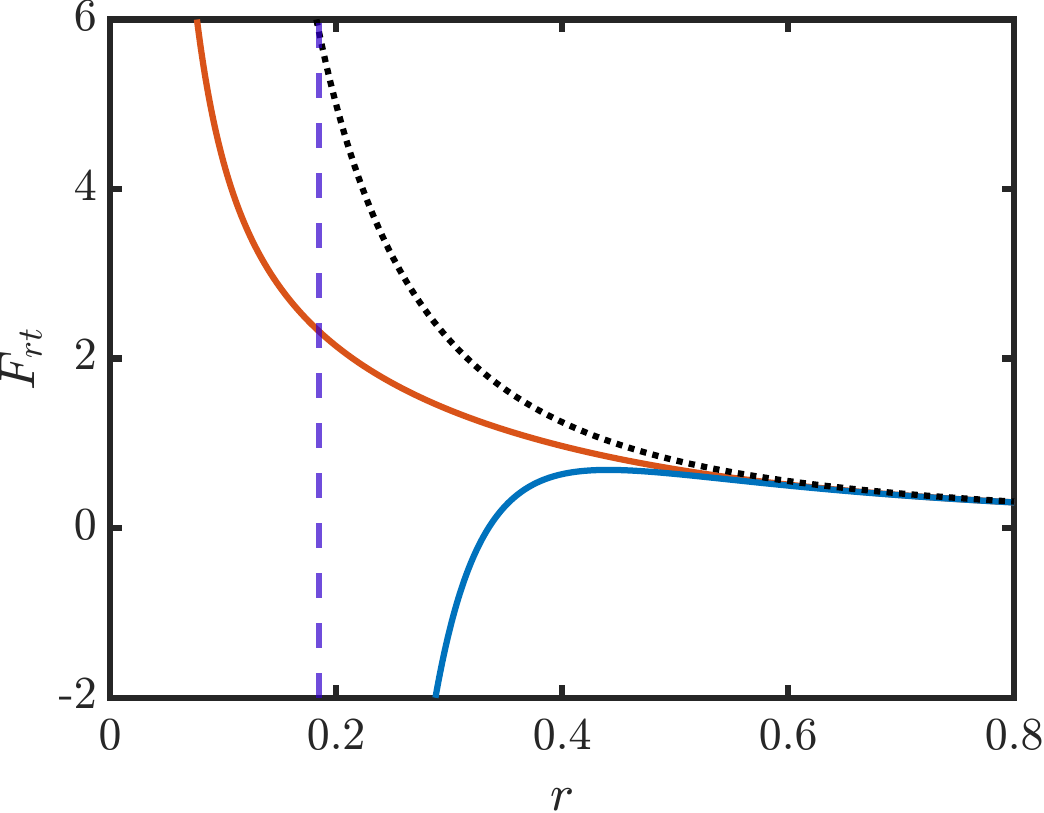}
    \end{minipage}
    
    \caption{Left: A plot of $F_{rt}$ against $r$ for $q = 0.2$, $p = 0.1$, $a = 0.05$ and $b = 0.037$, so that $2b - a  > 0$. Right: The same plot, but with $a = 0.08$ instead, so that $2b - a  < 0$. Both: The red curves are the numerical results, the blue are the perturbative results, the purple dashed line is the DEC violation threshold, and the black dotted curves are the analogous results for the linear (Maxwell) regime.}
    \label{NumericalSmallrPlots}
\end{figure}

Two examples are shown in Figure \ref{NumericalSmallrPlots}, where we see the failure of the perturbative solution for $F_{rt}$ below some radius $r$, compared to the true numerical solutions. In particular, we can see that for large $r$, the linear and the perturbative results are good approximations for the true results, but that for sufficiently small $r$, they both fail. For the left plot, the true numerical result is actually regular as $r \rightarrow 0$, as opposed to divergent with the linear and perturbative results. For the right plot, all curves are divergent as $r \rightarrow 0$, but the perturbative result has the wrong sign.
If this breakdown of the perturbative solution occurs outside the minimum event horizon, then this can invalidate the resulting turning point and phase diagram analysis.

\begin{figure}[t]
    \centering
        
    \begin{minipage}{0.47\textwidth}
        \includegraphics[width=\textwidth]{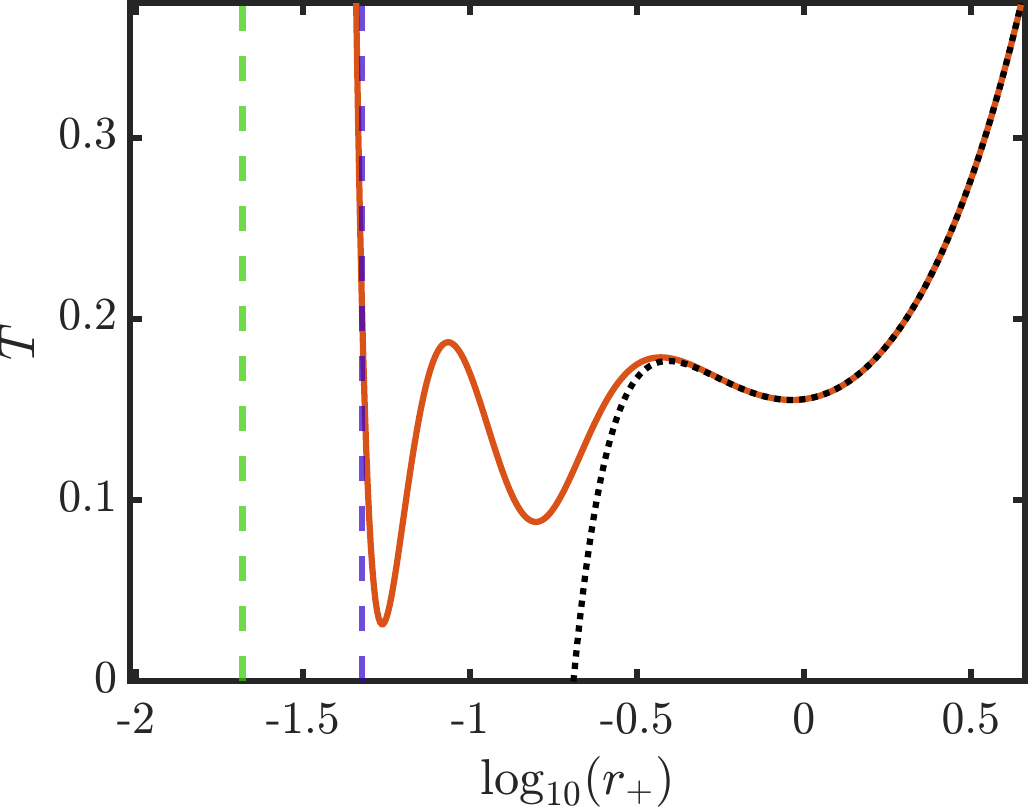}
    \end{minipage}
    \hspace{0.4cm}
    \begin{minipage}{0.47\textwidth}
        \includegraphics[width=\textwidth]{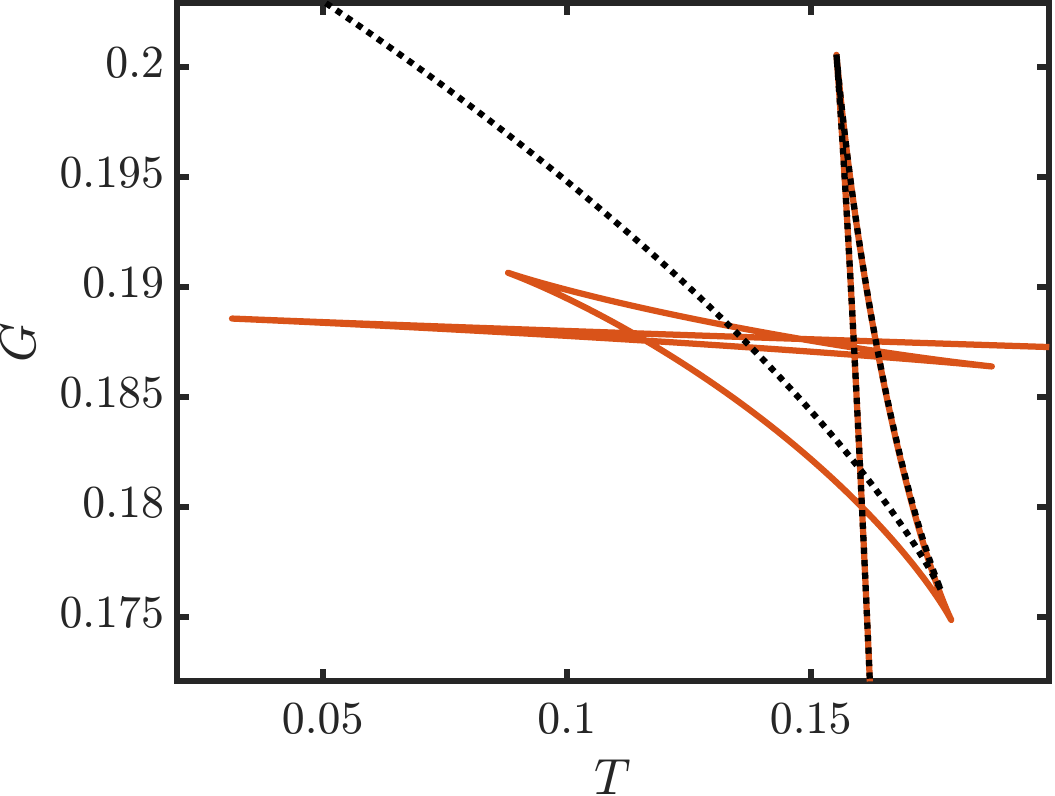}
    \end{minipage}
    
    \caption{Plots demonstrating a case with five turning points. Left: a plot of temperature $T$ with horizon radius $r_+$ for $q = 0.2$, $p = 0.059$, $a = 0.00037$, $b = 0.028$, $\Lambda = -1$. The green dashed line is the event horizon threshold, and the purple dashed line is the DEC violation threshold, which occurs just to the left of the smallest $r_+$ turning point. Right: a phase diagram of Gibbs free energy $G$ against temperature $T$ for the same choice of parameters. Both: The red curves are the numerical results and the black dotted curves are the analogous results for the linear RNAdS regime. This is a case with $2b - a > 0$, so the corresponding $F_{rt}$ will be similar to the left plot in Figure \ref{NumericalSmallrPlots}. Note that since $\Lambda = -1$, $r_+$ is dimensionless.}
    \label{NumericalPlotsEx}
\end{figure}

Up until this point, we have seen three turning points by a perturbative analysis (Figure \ref{LinearisedPlotsEx}), and four turning points from the special case $a = 0$ (Figure \ref{a0PlotsEx}). By working numerically, we can find up to \textit{five} turning points, with an example shown in Figure \ref{NumericalPlotsEx}.
As with Figure \ref{LinearisedPlotsEx}, despite the DEC violation boundary occurring near the smallest $r_+$ turning point, we will consider the turning points in Figure \ref{NumericalPlotsEx} to be valid, since they occur within the DEC `safe' region.

An extensive numerical search has revealed a maximum of five turning points available for spherically-symmetric black holes in the theory \eqref{action} for $a,b \geq 0$. Nonetheless, this is significantly more than the maximum of two available in the linear theory, discussed in Section \ref{CritRNAdSSec}.
Additional turning points have been reported previously in NLED theories \cite{Tavakoli:2022kmo}. However there, an expansion $\mathcal{L}(F)$ is considered with no $G$ dependence and no magnetic charge. To get as many as six turning points, terms up to $F^{7}$ seem to be required, with six new non-linear couplings in the theory. By including contributions from $G$, and magnetic charge $p$, we get a similar complexity of the phase space, but with little additional cost; only two new couplings $a,b$ in the theory.

We now provide some analytic demonstration that additional turning points at small $r$, beyond those identified in the perturbative analysis of Section \ref{PertSec}, can be expected within the theory \eqref{action}. To give the main point concisely, we will avoid special cases by considering $p,q \neq 0$ and $a,b \neq 0$. 

In investigating turning points, we will want an expression for the temperature $T$ in \eqref{GeneralThermoVariables} at small $r$. We start by extracting the dominant small $r$ behaviour of $F_{rt}$ from the cubic equation \eqref{Fcubic},
\begin{equation} \label{FrtSmallrCases}
   F_{rt} \sim
   \begin{cases}
     \frac{q}{4p^2 (2b-a)}\;  r^2  \, ,\quad & 2b-a > 0,\\
     \textrm{sign}(q) \sqrt{p^2(1-2\frac{b}{a})} \; r^{-2} \, , \quad & 2b-a < 0, \\
     (\frac{q}{4a})^{1/3}\; r^{-2/3} \, ,\quad & 2b-a = 0,
\end{cases}
\end{equation} 
as $r \rightarrow 0$. We will need to keep additional terms in $F_{rt}$ to get the small $r$ expansion of temperature $T$, but \eqref{FrtSmallrCases} gives the dominant behaviour for illustration. Note that $F_{rt}$ can have multiple roots for small $r$, particularly for $2b-a < 0$. The correct root must be chosen to match continuously with the $q/r^2$ approximate solution at large $r$, noting from the cubic \eqref{Fcubic} that $F_{rt}$ cannot cross through 0 for any $r$. The cases in \eqref{FrtSmallrCases} are consistent with the small $r$ behaviour shown in Figure \ref{NumericalSmallrPlots}. 

Next we can feed \eqref{FrtSmallrCases}, with sufficiently many extra terms, through to the temperature expression \eqref{GeneralThermoVariables}, finding,

\begin{equation} 
\label{TempSmallrCases}
   T \sim \frac{1}{4\pi}
   \begin{cases}
      \frac{2ap^4}{r_+^7} - \frac{p^2}{r_+^3} + \frac{1}{r_+} + \mathcal{O}(r_+)  \, , \quad & 2b-a > 0,\\
      \frac{8(1 - \frac{b}{a})bp^4}{r_+^7} - \frac{2(\frac{b}{a}p^2 + \sqrt{1 - \frac{2b}{a}}|pq|)}{r_+^3} + \frac{1}{r_+} + \mathcal{O}(r_+) \, , \quad & 2b-a < 0, \\
     \frac{2ap^4}{r_+^7} - \frac{p^2}{r_+^3} - \frac{3q}{2} (\frac{q}{4a})^{1/3} \frac{1}{r_+^{5/3}}+ \frac{1}{r_+} + \mathcal{O}(r_+^{1/3}) \, , \quad & 2b-a = 0.
\end{cases}
\end{equation} 

In each case, the dominant behaviour of $T$ at small $r$ has a positive $r_+^{-7}$ term and a negative $r_+^{-3}$ term.
Therefore a turning point $dT/dr_+ =0$ can occur from the competition of these terms, provided it is outside the minimum possible event horizon. 

In the linear theory, there is no such $r_+^{-7}$ term in the temperature \eqref{LinearTemp}, so a corresponding turning point will not appear.
With regards to the perturbative analysis of Section \ref{PertSec}, despite an $r_+^{-7}$ term appearing in the temperature \eqref{PertThermoVariables}, the competition of terms can occur below the threshold of validity \eqref{PertBreakdown} of the perturbative expansion, which would render it irrelevant in identifying a valid turning point.

To get an idea for this, we can calculate some numerical values for the example with five turning points shown in Figure \ref{NumericalPlotsEx}, which is a case with $2b-a > 0$. At the 10\% level, the linear theory breaks down below around $r = 0.302$, and the first order perturbative expansion breaks down below around $r = 0.230$, and we would not expect turning points in these ranges to be accurately captured by their respective analyses. From the analytic analysis of small $r$ above, the competition of $r_+^{-7}$ and $r_+^{-3}$ terms in \eqref{TempSmallrCases} gives the estimate $r_+ \approx (14ap^2/3)^{1/4} = 0.0495$, which is a reasonable approximation to the lowest turning point that occurs at $r_+ = 0.0547$.

\subsection{Weak field check}
\label{WeakFieldSec}

Since our action \eqref{action} is only a weak field expansion up to second order in $F, G$, it is reasonable to ask whether our solutions satisfy this notion.

\begin{figure}[ht]
\centering
\includegraphics[width=0.6\textwidth]{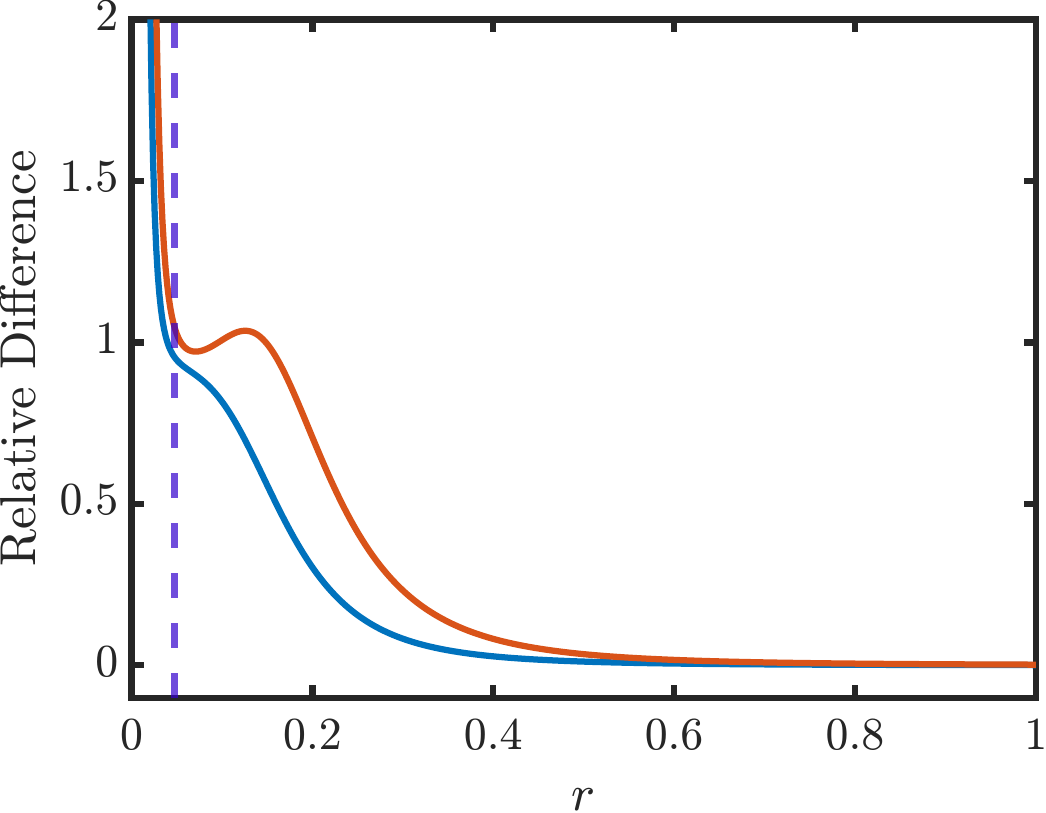}
\caption{A plot of the relative difference between the energy momentum components in the non-linear and linear solutions, in the case $q = 0.2$, $p = 0.059$, $a = 0.00037$ and $b = 0.028$, as in Figure \ref{NumericalPlotsEx}. The red curve shows the relative difference for $\rho$, and the blue curve for $p_{\theta}$. The purple dashed line is the DEC violation threshold, occurring at $r = 0.048$. $100\%$ relative difference is exceeded for $r \leq 0.149$, and $200\%$ is exceeded for $r \leq 0.028$, and since the lowest turning point occurs at $r_+ = 0.055$, this is a reasonably mild deviation above the former threshold. $\Lambda$ is irrelevant as only the matter contribution to the energy-momentum is considered. }
\label{WeakField_Ex}
 \end{figure}

Earlier, we used the dominant energy condition as one proxy for the weak field regime, however we can also compare the effect of the non-linear terms in the energy-momentum tensor against their linear counterparts.
We consider the relative difference between $T^{\text{non-linear}}_{\mu\nu}$ and $T^{\text{Maxwell}}_{\mu\nu}$, and demand that this is sufficiently small, preferably less than $100\%$, so that the non-linear terms are not having an excessive effect.
As we investigated in Section \ref{DECVioSec}, there are only two independent components of the energy-momentum, $\rho$ and $p_{\theta}$. 
An example of the relative differences of these quantities is shown in Figure \ref{WeakField_Ex}, matching the case of five turning points of Figure \ref{NumericalPlotsEx}.
Some $r_+$ exceed 100\% relative error within the range of interest, but certainly remain well within $200\%$ relative difference, and hence we consider this to be mild violation of this condition.
Note, Figure \ref{WeakField_Ex}, while presenting a specific example in parameter space, is representative of the relative size of the non-linear terms, as we have verified for a wide range of cases, including those we have presented throughout this work.

\subsection{Heat capacity and new phase transitions} \label{heat_cap_sec}

Thus far, we have seen examples of additional turning points arising in the black hole thermodynamics of the theory \eqref{action} compared to linear Maxwell theory. However, not all sections of the $(T, G)$ phase diagram will be thermodynamically stable.

To determine thermodynamic stability, the heat capacity $C = \frac{\partial M}{\partial T}$, with all other variables held fixed, can be calculated. 
For example, for a Schwarzschild black hole $C = -\frac{1}{8\pi T^2} < 0$. As the black hole evaporates, it undergoes a runaway process, the black hole loses mass and continues to heat up, evaporating faster. This suggests a thermodynamic instability.

\begin{figure}[t]
    \centering
        
    \begin{minipage}{0.47\textwidth}
        \includegraphics[width=\textwidth]{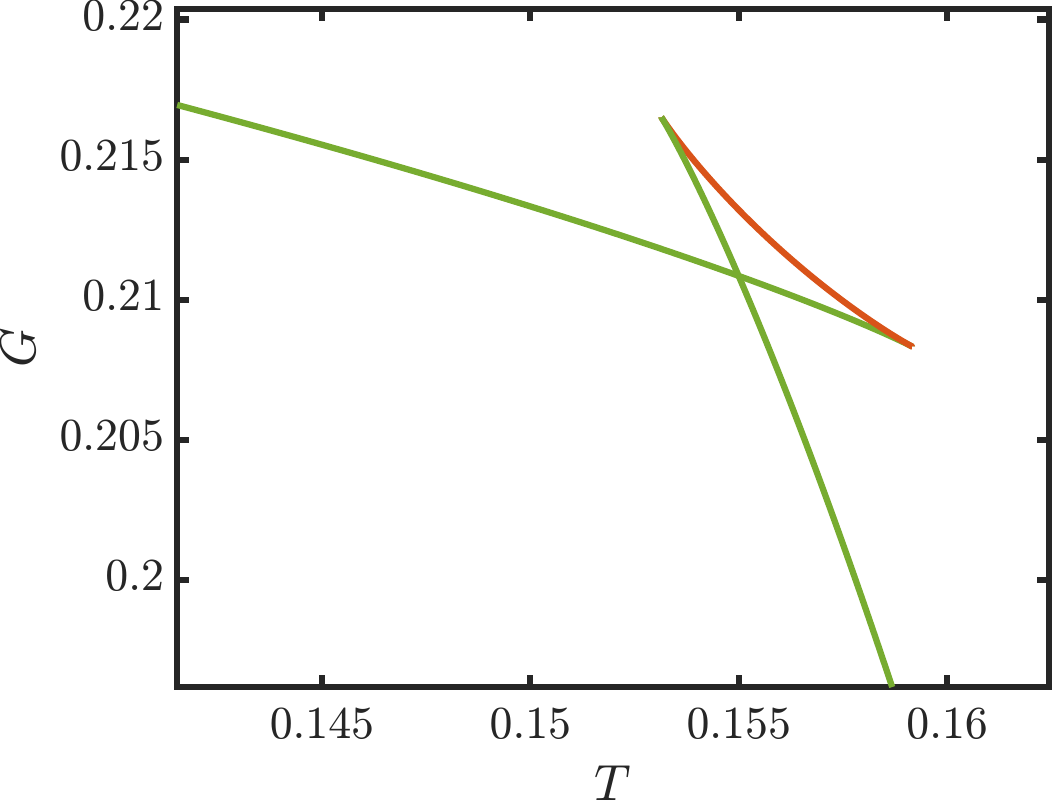}
    \end{minipage}
    \hspace{0.4cm}
    \begin{minipage}{0.47\textwidth}
        \includegraphics[width=\textwidth]{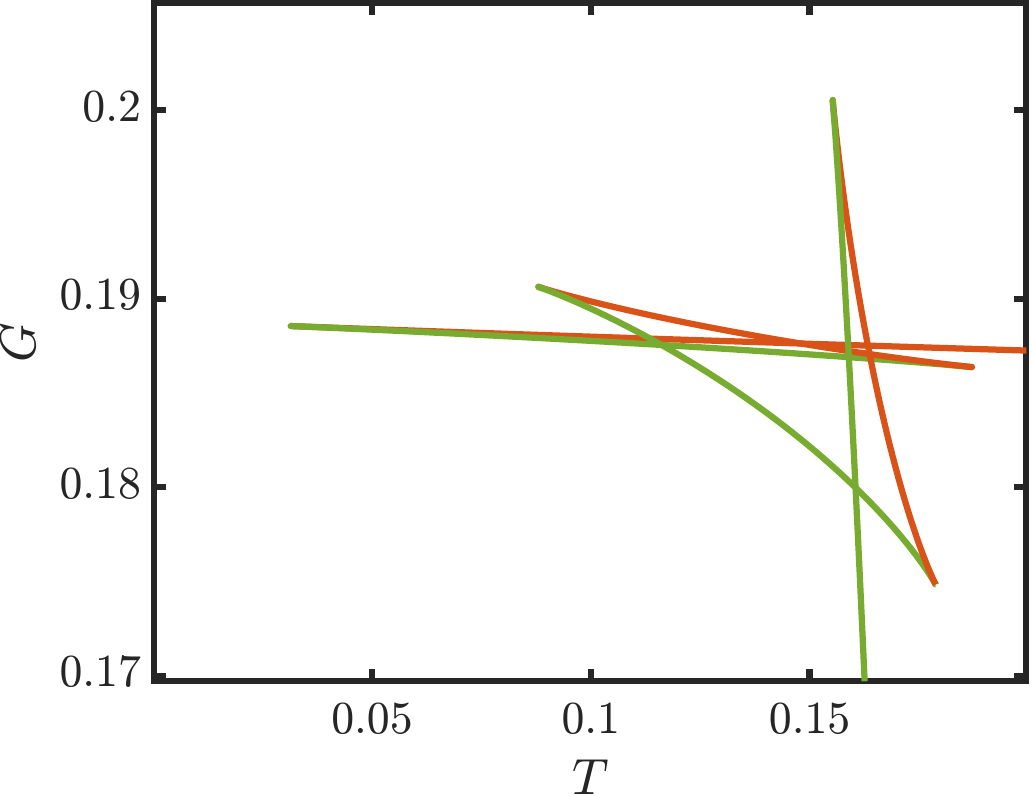}
    \end{minipage}
    
    \caption{Plots of phase diagrams displaying the sign of the heat capacity $C = \frac{\partial M}{\partial T}$. Sections with $C \geq 0$ are shown in green (stable), and sections with $C < 0$ are shown in red (unstable).
    Left: RNAdS with $q = 0.25$, $p = 0$, $a = 0$, $b = 0$ and $\Lambda = -1$, matching a curve in Figure \ref{RNAdSPhaseEx}. Right: a non-linear case, with $q = 0.2$, $p = 0.059$, $a = 0.00037$, $b = 0.028$ and $\Lambda = -1$, matching Figure \ref{NumericalPlotsEx}.
    }
    \label{heat_examples}
\end{figure}

The heat capacity $C$ can be calculated for each $r_+$, via $M(r_+)$ and $T(r_+)$. We can hence use $C < 0$ as an indication of thermodynamic instability of a black hole configuration with horizon $r_+$ \cite{Altamirano:2013ane}.
We show examples of this for Maxwell theory and for an NLED solution in Figure \ref{heat_examples}. The heat capacity tends to change sign at a turning point, but it is possible to have the same sign on both sides, typically in second-order (or higher) phase transitions.

Now that we have explored the turning point structure in NLED, and clarified thermodynamic stability, we can ask if the non-linear terms in the action \eqref{action} induce any non-trivial phase transitions.
Firstly, we will consider the phase transition structure of our existing phase diagrams with 3, 4 and 5 turning points.
Figure \ref{LinearisedPlotsEx} shows a first-order phase transition, as the temperature is initially decreased. However, as the temperature is lowered further, there is another first-order phase transition, at a critical temperature below which no black hole is thermodynamically preferred, analogous to the Hawking-Page transition in AdS \cite{hawking-page1983}.
Figure \ref{a0PlotsEx} shows two first order phase transitions, as the temperature is decreased.
Figure \ref{NumericalPlotsEx} similarly shows two first order phase transitions, and then a final first-order phase transition, as in Hawking-Page.
In each of these cases, the initial first-order phase transition is analogous to that which occurs in the swallowtail of RNAdS, but then the non-linearities take effect at lower horizon radius $r_+$ to offer new phase structure.

\begin{figure}[t]
    \centering
        
    \begin{minipage}{0.47\textwidth}
        \includegraphics[width=\textwidth]{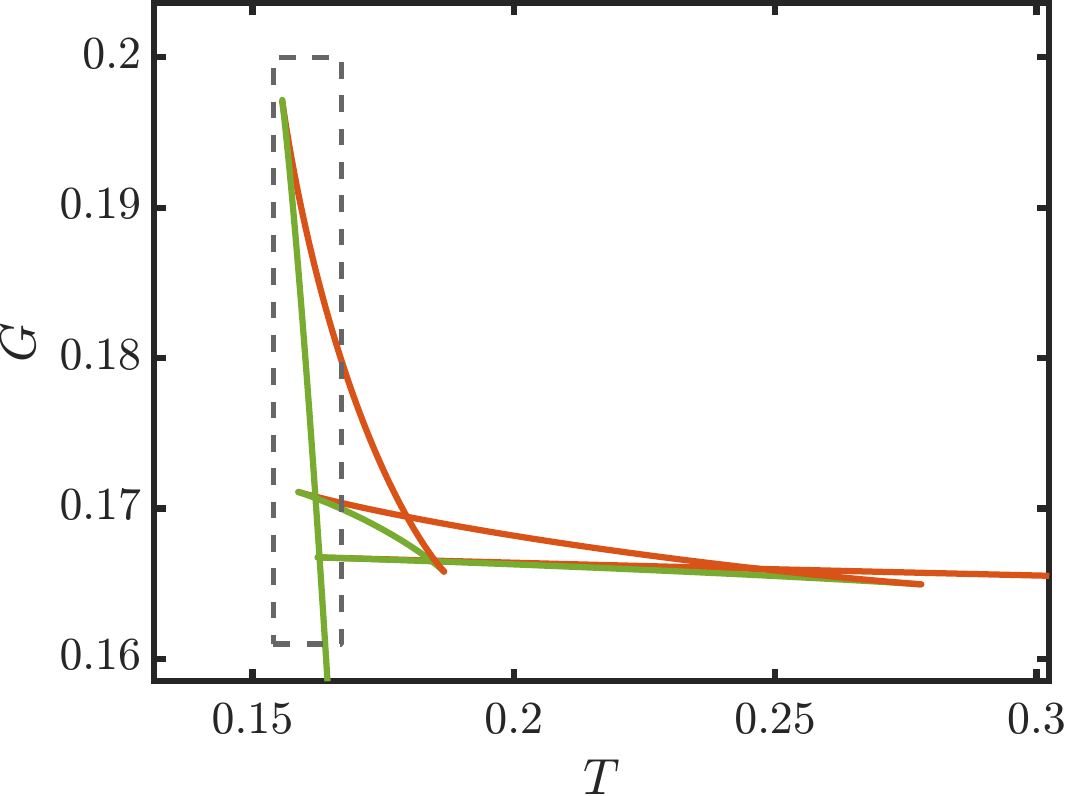}
    \end{minipage}
    \hspace{0.4cm}
    \begin{minipage}{0.47\textwidth}
        \includegraphics[width=\textwidth]{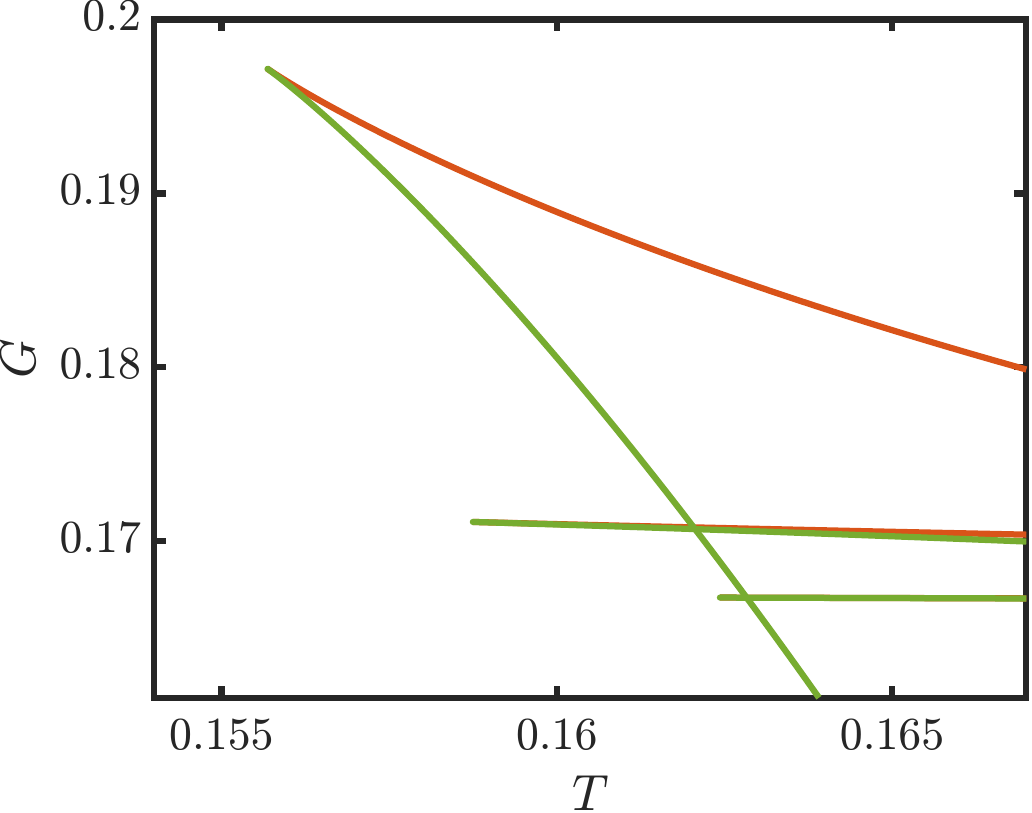}
    \end{minipage}
    
    \caption{A plot showing multiple reentrant phase transitions, for $\Lambda = -1$, $q = 0.189$, $p = 0.06$, $a = 0.0004463$ and $b = 0.035$. The curves are green for $C \geq 0$, and red for $C < 0$, indicating the thermodynamic stability. The left plot shows the overall phase diagram, and the right plot shows a zoomed window with a different aspect ratio (highlighted as a grey dashed region on the left plot) into the lower temperature region, where the multiple reentrant phase transitions occur. }
    \label{Multi_ReEnt_Ex}
\end{figure}

Furthermore, reentrant phase transitions can occur when the solution transitions from one phase, to another, and then back again. These do not occur for RNAdS in $d = 4$, but they have been noted previously in NLED theories \cite{Dehyadegari:2017hvd}. Here we give an example of a solution with multiple reentrant phase transitions, between three different branches, shown in Figure \ref{Multi_ReEnt_Ex}. Here, the thermodynamically-preferred solution transitions from large black holes (LBH), to an intermediate black hole branch, back to LBH, back to another intermediate branch, and then finally back to the LBH branch. Some of these phase transitions are zeroth-order, where $G$ itself is discontinuous as the transitions occur. Single reentrant phase transitions in black hole thermodynamics have been noted in NLED theories \cite{Dehyadegari:2017hvd}, and multiple reentrant phase transitions have been noted previously in Lovelock theories \cite{Frassino:2014pha}, albeit cycling between the same pair of phases. 

Cases with superfluid-like phase transitions have also been found in Lovelock thermodynamics \cite{Hennigar:2016xwd}, so we may expect this to exist in the NLED system \eqref{action}, but we leave this to future investigation.

\section{Discussion}

One of the core challenges in modern theoretical physics is understanding the precise nature of the quantum vacuum \cite{Weinberg:1988cp}. Non-linear electrodynamics can alter this structure in strong-field regimes, predicting phenomena such as vacuum birefringence and light-by-light scattering. NLED can also have a dramatic effect on black hole spacetimes, of which the regularisation of singularities in black hole solutions \cite{ayon1999} is the most startling.
The non-linear gauge terms appearing in NLED arise naturally in string-inspired theories \cite{Fradkin:1985qd} and consequently, studying black holes within NLED frameworks also advances our understanding of black hole physics in string theory contexts.
 
In this work, we have presented a systematic analysis of the impact of weak NLED fields on black holes and their thermodynamics. We have derived the first law of thermodynamics and the free energy, allowing an exploration of the rich thermodynamic phase structure that NLED facilitates.
Analytic investigations have provided intuitive understanding of new features that black holes can acquire in NLED theories, such as a lack of extremal limit.
With a numerical approach, we have found examples with up to five turning points, three more than the linear Maxwell theory. Moreover, we have noted examples of multiple reentrant phase transitions, which have been seen previously in modified gravity \cite{Frassino:2014pha}.

We conclude by proposing potential extensions and unresolved questions that could be addressed in future studies.
In Section \ref{ExtremalJumpSec}, we found black hole solutions with stationary points in the gravitational potential outside the horizon. This curious feature warrants further investigation, possibly leading to interesting phenomena, such as stable light rings \cite{Liu:2019rib} or a more complex quasi-normal mode structure.
As we suggested in Section \ref{DECVioSec}, it may be possible to evade the DEC constraint \eqref{abconstraint} with $a \leq 0$ or $b \leq 0$. This could lead to further interesting scenarios as the limit of energy conditions are tested.
In \cite{Tavakoli:2022kmo}, Tavakoli \textit{et al}.\ found multi-critical points, where multiple branches coincide at one position on the temperature/free-energy phase diagram. In order to achieve such points, they required an NLED theory with high powers of $F$, without any consideration of $G$. Since we have a similarly complex phase structure in our results, it is likely that these multi-critical point scenarios will also occur in our NLED theory \eqref{action} while only utilising a second order weak field expansion.
Having appreciated the additional complexity that second order has to offer, one could take a step further and consider additional effects at third order, including the parity and time-reversal conserving terms $F^3$ and $FG^2$ in the action \eqref{action}.
Furthermore, one could also consider parity and time-reversal violating terms, such as $FG$ at second order, 
or additional terms could be included to capture NLED theories such as ModMax \eqref{ModMaxLag}.
The diverse shapes revealed in our phase diagrams not only highlight the complexity of the system but also open a compelling pathway: through topological thermodynamics, distinct catastrophe topologies can be identified and characterised, potentially unveiling deeper insights into the nature of NLED black holes.

\acknowledgments

The authors thank Ansh Gupta for correcting some typos in an earlier version of this paper, and an anonymous referee for useful suggestions. LC is supported by King’s College London through an NMES funded studentship. RG is supported in part by the STFC grant (ST/X000753/1) and in part by the Perimeter Institute for Theoretical Physics. CRV acknowledges support from the Secretaría de Educación, Ciencia, Tecnología e Innovación de la Ciudad de México (SECTEI) of Mexico City. Research at Perimeter Institute is supported in part by the Government of Canada through the Department of Innovation, Science and Economic Development and by the Province of Ontario through the Ministry of Colleges and Universities.

\appendix

\section{Derivations from Section \ref{ExtremalJumpSec}} \label{ExtremalJumpDerivations}

In this appendix, we derive the conditions on the $q = 0$ solution, stated in Section \ref{ExtremalJumpSec}, regarding extremal limits and jumping horizons.

First, on extremal limits, looking for a locally extremal black hole horizon suggests we need $f(r_+) = 0$ at a local minimum or inflection, so $ f'(r_+) = 0$ and $f''(r_+) \geq 0$. Eliminating the mass $m$ gives conditions,
\begin{equation}
    \begin{aligned}
    r_+^6 - p^2 r_+^4 + 2ap^4 & = 0, \\
    r_+^4 & \geq 6ap^2.
\end{aligned}
\end{equation}
Let $g(x) = x^3 - p^2 x^2 +2ap^4$, so that the former is a cubic in $x = r_+^2$. Since $x = 0$ gives a local maximum with $g(0) = 2ap^4 > 0$, and we need a solution with $x > 0$, there must be a local minimum with $x^* > 0$ and $g(x^*) \leq 0$. 
The other stationary point occurs at $x^* = \frac{2}{3}p^2$, so we need $g(\frac{2}{3}p^2) \leq 0$, leading to the condition
\begin{align}
    p^2 \geq \frac{27}{2}a.
\end{align}
It can be shown that this condition is necessary and sufficient for a local extremal limit, and that the resulting mass $m$ is positive.

Next we derive the range of parameters for which a jump in horizon radius occurs as the mass $m$ is varied.  
As noted in the main text, to have a jump in $r_+$ as we vary $m$, we need $m(r_+)$ to be a non-monotonic function, but in some range of $r_+$ that does not violate DEC (see Figure \ref{mass_jump}). Hence $m(r_+)$ needs a local minimum, say at $r_+ = r_M$.
If the DEC violation threshold is $r_+ = r_D \equiv (4ap^2)^{1/4}$, then the presence of a jump leads to the condition $m(r_D) \leq m(r_M)$.
To find the range of parameters that satisfy these conditions, we first find the case that saturates this inequality. This is equivalent to requiring a minimum in $h(r_+) \equiv m(r_+) - m(r_{D})$, where simultaneously $h(r_+) = 0$.
Factoring out $(r_+ - r_{D})$, and requiring a stationary point in what remains,
we have,
\begin{equation}
    \begin{aligned}
        r_+^5 - \frac{9p^2}{10(4ap^2)^{1/4}} r_+^4 + \frac{p^2}{10} r_+^3 + \frac{p^2 (4ap^2)^{1/4}}{10} r_+^2 + \frac{p^2 (4ap^2)^{1/2}}{10}r_+ + \frac{p^2 (4ap^2)^{3/4}}{10} = 0, \\
        5r_+^4 - \frac{18p^2}{5(4ap^2)^{1/4}} r_+^3 + \frac{3p^2}{10} r_+^2 + \frac{p^2 (4ap^2)^{1/4}}{5} r_+ + \frac{p^2 (4ap^2)^{1/2}}{10} = 0.
    \end{aligned}
\end{equation}
Solving simultaneously and eliminating $r_+$ gives the polynomial equation,
\begin{align} \label{PolyJumping}
    132111 s^4 - 92816 s^3 -124248 s^2 - 173248 s - 250000 = 0,
\end{align} 
in terms of $s = \sqrt{p^2/4a}$.
This polynomial only has one positive root, and it corresponds to a minimum in $m(r_+)$, with $m(r_+) > 0$, provided $p^2 > 27a/2 $.  Hence the range of $p^2$ for a jump in horizon radius is,
\begin{align}
   \frac{27}{2}a < p^2 \leq \eta \, a,
\end{align}
where $s = \sqrt{\eta}/2$ is the only positive root to the polynomial equation \eqref{PolyJumping}. $\eta$ can be reduced to a lengthy expression in radicals, but it takes the value $\eta \approx 13.987$.

\bibliography{bib.bib}
\bibliographystyle{JHEP}

\end{document}